\documentclass[aps,pra,twocolumn,groupedaddress,amsmath,amssymb]{revtex4}
\usepackage{graphicx}
\usepackage{natbib}

\begin{document}
\newcommand{\Fig}{Fig.~}
\newcommand{\Figs}{Figs.~}
\newcommand{\Figstart}{Figure~}
\newcommand{\Rb}{$^{87}$Rb }
\newcommand{\St}{Stokes }
\newcommand{\aS}{probe }
\newcommand{\fxm}[1]{{\bf !---!FIXME #1 !---!}}
\newcommand{\etal}{\textit{et al.~}}
\newcommand{\hideit}[1]{}
\newcommand{\notforpaper}[1]{}

\author{Eugeniy E.\ Mikhailov}
    \email{evmik@leona.physics.tamu.edu}
\author{Irina Novikova}
\author{Yuri V.\ Rostovtsev}
\author{George R.\ Welch}

\affiliation{
    Department of Physics and Institute for Quantum Studies,
    Texas A\&M University,
    College Station, Texas 77843-4242
}

\title{ 
    Buffer-gas induced absorption resonances in Rb vapor
} 

\begin{abstract} 
We observe transformation of the electromagnetically induced
transparency (EIT) resonance into the absorption resonance in a
$\Lambda$ interaction configuration in a cell filled with $^{87}$Rb and
a buffer gas.  This transformation occurs as a one-photon detuning
of the coupling fields is varied from the atomic transition.  No such
absorption resonance is found in the absence of a buffer gas. The
width of the absorption resonance is several times smaller than
the width of the EIT resonance, and the changes of absorption near
these resonances are about the same. Similar absorption resonances
are detected in the Hanle configuration in a buffered cell.
\end{abstract} 

\date{\today}
\maketitle

\section{Introduction} 

    The coherent interaction of atoms with electromagnetic
fields has attracted increasing attention recently in studies of
nonlinear and quantum optics as well as spectroscopy and precision
metrology. 
Under the combined action of several resonant laser fields, atoms
are optically pumped into a coherent superposition of the
ground-state (hyperfine or magnetic sublevels) which is decoupled
from the original electromagnetic fields.  That is, the atoms are
in a so-called ``dark'' state.  Such a medium possesses some
unique optical properties, for example, coherent population
trapping (CPT)~\cite{arimondo'76,agapyev'93, arimondo'96po},
cancellation of absorption due to electromagetically induced
transparency (EIT)~\cite{scullybook, harris'97pt, marangos'98},
and steep nonlinear dispersion~\cite{scullybook, harris'92,
schmidt'96, renzoni'00}. The characteristic spectral width of
features occurring due to these nonlinear effects is determined by
the inverse lifetime of an atom in the coherent superposition of
ground states. Since radiative transitions are usually forbidden
between these states, the coherence can be preserved for a long
time, and in atomic cells its lifetime is usually determined by
the interaction time of the atom with the laser beams
~\cite{arimondo'96pra, graf'95}.

    The addition of a buffer gas (inert  gases, $\mathrm{N}_2$,
$\mathrm{CO}_2$, $\mathrm{CH}_4$, etc.)\ to the atomic vapor is a
common method for obtaining narrow EIT resonances.  Because of
the extremely low spin-exchange cross-section, the collisions
between rubidium and buffer gas atoms or molecules do not
destroy the quantum coherence of the internal states of the
atoms, but effectively prolong the time they stay inside the
laser beam(s).  The processes of decoherence and redistribution
of atomic population have been extensively studied in optical
pumping experiments~\cite{bernheim_book, vanier98, happer'72}.
Substantial narrowing of the dark resonance linewidth is
reported in~\cite{brandt'97, wynands'98, helm'00, helm'01}.

\begin{figure} 
\includegraphics[angle=0, width=1.00\columnwidth]{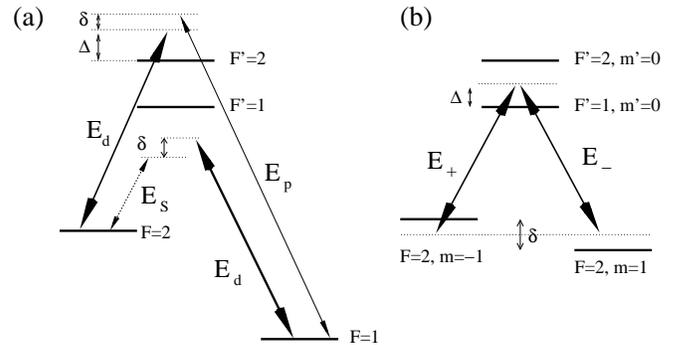}
\caption{
    (a) Three-level interaction scheme of three laser fields
    with $^{87}$Rb atoms: the long-lived coherence is created on
    hyperfine ground-state sublevels with strong driving
    field $\mathrm{E_{d}}$ and weak probe (anti-Stokes)
    field $\mathrm{E_{p}}$; the probe and Stokes field
    $\mathrm{E_{s}}$ are generated by electro-optic
    modulation.
    (b) Hanle configuration: the ground-state coherence
    is created on magnetic sublevels with two circularly
    polarized components $\mathrm{E}_{\pm}$ of a monochromatic
    linearly polarized laser field.
    In both cases $\Delta$ is the one-photon detuning
    of the laser(s) from atomic resonance, and $\delta$
    is the two-photon detuning due to frequency mismatch
    in case (a) or an external magnetic field in case (b).
\label{levels.fig}
}
\end{figure}

In this paper we study the transformation of the transmission peaks 
corresponding to EIT into enhanced absorption peaks for proper laser
detuning. We present an extensive experimental and theoretical
analysis of this effect,  previously reported in 
~\cite{mikhailov04josab,mikhailov04praprep}. We observe narrow enhanced
absorption resonances  in two experimental arrangements: in a
bi-chrimatic 
configuration, where strong and weak laser fields form a $\Lambda$
scheme on two ground-state hyperfine sublevels
(Fig.~\ref{levels.fig}a), and in the degenerate Hanle configuration
(Fig.~\ref{levels.fig}b). In both situation the EIT resonance is observed
for the laser field(s) tuned to the atomic transition(s) due to formation of 
a non-interacting ground-state atomic coherence; in both cases the lineshape
of this transmission peak persists for any value of the one-photon
detuning unless a buffer gas is added to the Rb vapor.

Before proceeding into the details of the experiment let us
highlight the distinctions between the effect described below, and
similarly looking coherent effects:

(i) A transformation of enhanced transmission to enhanced absorption
CPT resonance has been reported by Affolderbach \textit{et
al.}~\cite{affolderbach'02} in their experiments with a bichromatic standing
wave in hot atomic vapor. Because of large Doppler
broadening, moving atoms effectively interact with a double-$\Lambda$
level configuration, which may result in either suppression or enhancement of
absorption, depending on the relative phases of the driving fields. This
explanation, however, cannot be applied to the present experimental data,
since our experiments are carried out with running waves.

 (ii) Narrow absorption resonances in alkali vapors are also the
 manifestation
of Electromagnetically Induced Absorption (EIA)~\cite{akulshin'98,
lazema'99, lipsich'00, taichenachev'00jetp, taichenachev'00pra,
kwon'01, ye'02}, or closely related enhanced absorption Hanle
effect~\cite{dancheva'00, alzetta'01, arimondo'01job,
arimondo'01pra, andreeva'02}.  In those experiments, a narrow
absorption peak appears for two laser fields of close frequencies
interacting with a quasi-degenerate two-level system.  However, it
is a general requirement in both cases that the degeneracy of the
ground state must be lower than that of the excited state, i.e. $F
< F'$, which prohibits any dark state formation. Narrow EIA
resonances in this case are due to the spontaneous coherence
transfer from the excited states of the
atoms~\cite{taichenachev'00jetp, taichenachev'00pra,
Failache2003pra}.  In addition, the experimental arrangements for
traditional EIA experiments involve laser fields resonant with the
corresponding atomic transitions, whereas the narrow absorption
resonances discussed in this paper appear for far-detuned laser
fields.

(iii) Finally, the effect described below cannot be attributed to pressure
induced extra resonances (PIER)(for a review of the collision-induced coherent
effects see ~\cite{agarwal'91aamop}). In PIER the extra resonance is connected
to a dressed state population that is non-vanishing only in the presence of
collisions. A relatively low buffer gas pressure used in our experiment does not
allow the realization of PEIR.

    This paper is organized as follows.  In the next
Section we describe the experimental apparatus and measurement
technique.  In Section III we present the experimental study of
these resonances in the three-level $\Lambda$ scheme based on the
hyperfine coherence in atomic cells with different amounts of a
buffer gas.  Theoretical analysis in given in Section IV\@.
Resonant four-wave mixing and the lineshape of the associated
resonances in a Stokes field are discussed in Section V\@. A brief
analysis of the coherent resonances' width is given in Section
VI\@. In Section VII, we present the experimental results on the
enhanced absorption resonances observed in the
Hanle-configuration, together with a qualitative discussion of the
their origin.  A brief summary of the work appears in the final
Section.


\section{Experimental setup} 

\begin{figure} 
\includegraphics[angle=0, width=1.00\columnwidth] {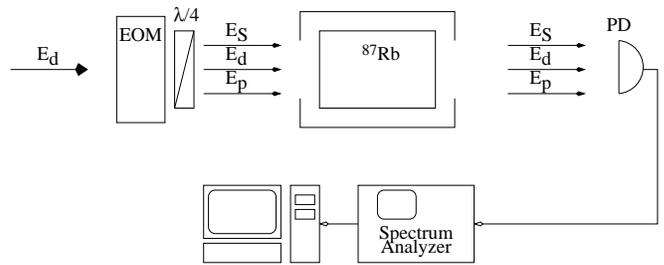}
\caption{
    Schematic of the experimental setup.
\label{simple_setup}
}
\end{figure}

    A schematic of the experimental setup is shown in
Fig.~\ref{simple_setup}.  An external cavity diode laser is tuned
to the $5S_{1/2} \rightarrow 5P_{1/2}$ ($D_1$) transition of $^{87}$Rb.
A probe field $\mathrm{E_p}$ (and an additional Stokes field
$\mathrm{E_S}$) are produced by an electro-optic modulator (EOM)
driven by a stable narrow-band tunable microwave generator
operating at 6.835~GHz to match the $^{87}$Rb ground-state hyperfine
transition
frequency.
The probe and Stokes fields have equal intensities of
approximately $10\%$ of that of the drive field. After the EOM,
all fields pass through a single-mode optical fiber to create a
clean spatial mode with a Gaussian radial intensity distribution and to
increase the diameter of the output beam to 7~mm.  The fields are
circularly polarized with a quarter-wave plate placed after the
fiber.

    In this experiment we use several glass cells filled
with isotopically enhanced $^{87}$Rb and various pressures and types
(Ne, Kr) of a buffer gas.  Each cell is placed inside a 3-layer
magnetic shield to screen the laboratory magnetic field from the
system and is heated to $60^o$ C to increase the density of the
$^{87}$Rb vapor to approximately $2.5\times 10^{11}~\mathrm{cm}^{-3}$.
After traversing the cell, all three fields are mixed on a fast
photodiode with an additional field shifted by 60~MHz with respect
to the driving field. The resulting photo-current is measured with
a spectrum analyzer to separate the transmission signals of the
probe and Stokes components (see Fig.~\ref{simple_setup}).  This
detection scheme has been described
in~\cite{kash99,mikhailov2002}.

    For our studies in the Hanle configuration, the
electro-optical modulator is removed so that there is only
one linearly polarized monochromatic electromagnetic field
propagating through the cell.  The transmitted intensity
is recorded directly from a photodetector.  A longitudinal
magnetic field for shifting the Zeeman sublevels is created
by a solenoid mounted inside the inner magnetic shield.

\section{Enhanced absorption due to a buffer gas in hyperfine $\Lambda$ scheme}

\begin{figure*} 
\includegraphics[angle=0,
    width=1.50\columnwidth]
{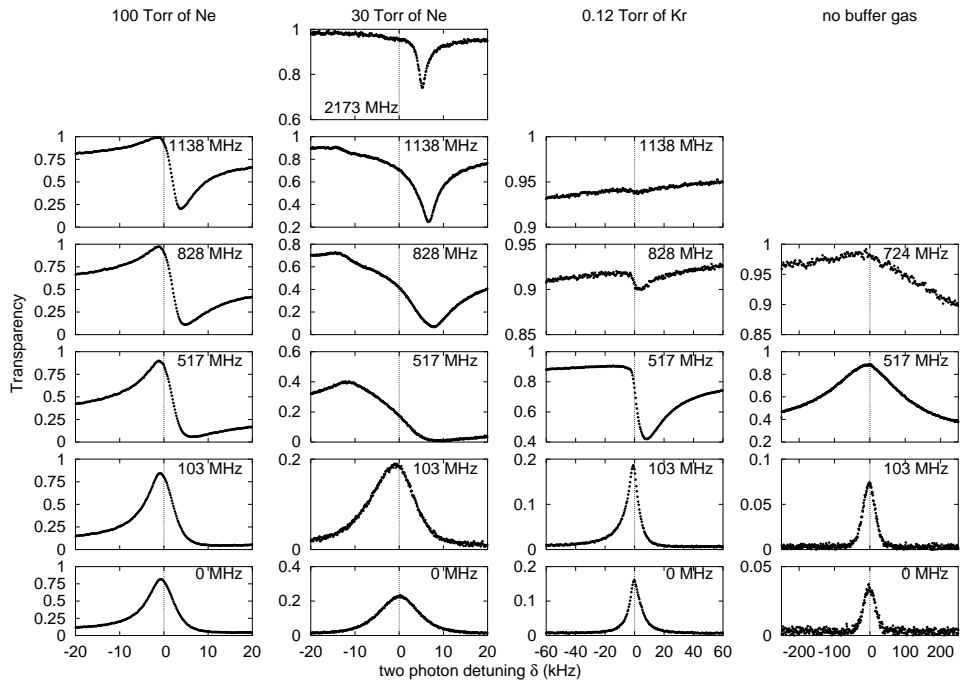}
\caption{
    Transmission of the probe field as a function of
    two-photon detuning $\delta$ for various one-photon
    detunings $\Delta$.   The values of $\Delta$ are shown
    on the upper right corner of each graph.  These data are
    recorded in the presence of (columns, left to right)
    100~Torr of Ne, $30$~Torr of Ne, $0.12$~Torr of Kr,
    and no buffer gas.  All signals are normalized to 
    the non-absorbed transmission.  The asymmetry of the
    resonance curves for $\Delta = 828$ and $1138$~MHz in
    the cell with $0.12$~Torr Kr, and for $\Delta = 517$
    and $740$~MHz in the vacuum cell are due to the slope
    of the one-photon absorption contour. The intensities of 
    the laser fields are $I_d\simeq 1.20\mathrm{mW/cm^2}$ 
    and $I_p\simeq 0.13\mathrm{mW/cm^2}$.
    \label{eit_shape_example}
}
\end{figure*}

    We start with the drive field resonant with
$5S_{1/2} F=2 \rightarrow 5P_{1/2} F'=2$ transition;
simultaneously, the probe field couples the ground-state $5S_{1/2}
F=1$ level to the same excited state.  In that resonant $\Lambda$
system we observe expected narrow EIT resonance in the probe field
transmission due to efficient optical pumping into non-interacting
dark state. The bottom row of Fig.~\ref{eit_shape_example} shows
the observed resonances for the cells with different amount of a
buffer gas. It is easy to see that all EIT peaks are nearly symmetric,
which is in agreement with theoretical
predictions~\cite{javan'02,rost'02,kuznetsova'02, lee'03} and our
numerical simulations.  The main difference between different
cells is the narrowing the EIT width resulting from the increased
interaction time for higher buffer gas pressure.

As we tune the laser away from the atomic transition (while
maintaining near-zero two photon detuning) the shape of the
transmission resonance changes depending on the presence of a
buffer gas in a cell. For the vacuum cell (the rightmost column in
Fig.~\ref{eit_shape_example}) the resonance stays symmetric while
its width increases rapidly with one-photon detuning $\Delta$ due
to less efficient density narrowing.

    However, the behavior is very different in cells
with a buffer gas.  As $\Delta$ increases, the EIT resonance
becomes asymmetric, and then gradually transforms into a narrow
\textit{absorption} resonance. Let us emphasize here some of the
important properties of these resonances.  For example, the cell
with $30$~Torr of Ne (Fig.~\ref{eit_shape_example}) shows that the
amplitude of the enhanced absorption peak observed for large
detuning ($\Delta\approx 1$~--~2~GHz) is comparable to, and
sometimes larger than, the amplitude of the EIT peak at
$\Delta=0$, and its width is narrower.  Second, these narrow
absorption resonances are observed for the laser detuning
exceeding the Doppler-broadened one-photon resonant absorption
width ($\Delta>1$~GHz). Thus, the enhanced absorption peaks
appears on top of $100\%$ transmission of the probe field.

   The asymmetry
in the EIT resonances for nonzero one-photon detuning of the laser
fields has been observed by Levi \textit{et
al.}~\cite{Levi2000epjd} in maser emission in the CPT process as well as by
Knappe \textit{et
al.}~\cite{knappe2003} .
However, the reported modification of the resonance lineshape is
significantly weaker. Knappe \textit{et al.} also suggested an
empirical expression for the resonance line-shape:
\begin{equation}
    \label{res_shape_fit}
    f(\delta)=\tilde \gamma \frac{A \tilde{\gamma} + B (\delta-\delta_0)}
    {\tilde \gamma^2 + (\delta-\delta_0)^2} + C
\end{equation}
Here $\delta$ is a two-photon (Raman) detuning, $\tilde\gamma$ is
an effective width of the coherent resonance, and $A$, $B$, and
$C$ are fitting parameters which are functions of the one-photon
detuning $\Delta$.  We introduce a shift $\delta_0$ of the
resonance position from the exact Raman condition to reproduce the
experimental data.  One can see that this expression consists of
symmetric and anti-symmetric Lorentzian functions of $\delta$ with
amplitudes $A$ and $B$ respectively. Parameter $C$ reflects the
residual absorption of the electromagnetic field determined by
incoherent processes. Taichenachev \textit{et
al.}~\cite{taichen2003} derived the analytical expressions for these
coefficients in the limit of weak interaction fields
(perturbation approach). The case of the strong (drive) field is discussed
in the next Section.

    It is convenient to write the coefficients A and B in the
following form:
\begin{equation}
A=D \cos(\phi)\,, \quad B=D \sin(\phi) \, ,
\end{equation}
so that the Eq.~(\ref{res_shape_fit}) can be written as:
\begin{equation} \label{res_shape_fit_D}
f(\delta)=\Re\left\{ D(\Delta)
e^{i\phi(\Delta)}\frac{\tilde \gamma}{\tilde \gamma +
i(\delta-\delta_0)}\right\} +C\,.
\end{equation}
In this case, the parameter $D$ characterizes the amplitude
of the resonance, and the angle $\phi$  expresses the
ratio between the symmetric and anti-symmetric components in
Eq.~(\ref{res_shape_fit}).  For example, $\phi=0$ represents
a symmetric peak of enhanced transmission, $\phi = \pi$
corresponds to a symmetric peak of enhanced absorption,
and $\phi = \pm \pi/2$ corresponds to a pure dispersion-like
lineshape.

\begin{figure} 
\includegraphics[angle=0,
    width=1.00\columnwidth]
    {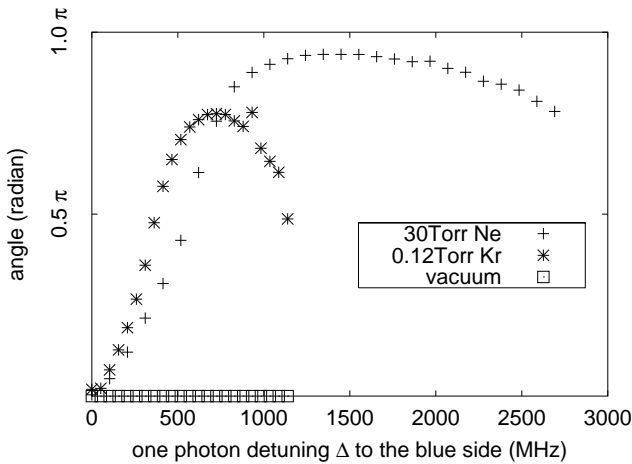}
\caption{
    Angle $\phi$ of two photon resonance for $^{87}$Rb cells with
    different amount of a buffer gas.
    \label{angle_aS.fig}
}
\end{figure}

\begin{figure} 
\includegraphics[angle=0,
    width=1.00\columnwidth]
    {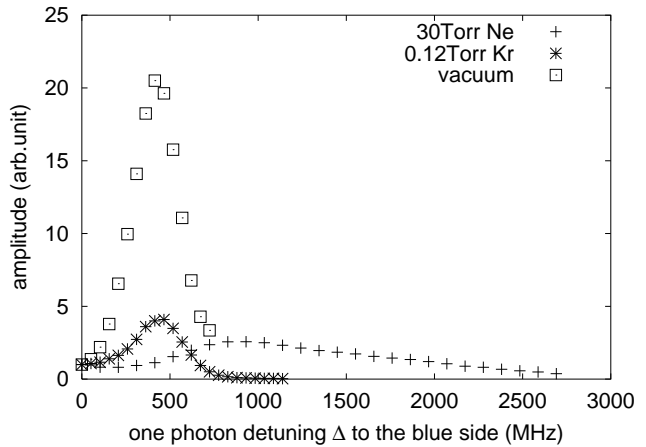}
    \caption{
    Amplitude $D$ of the two-photon resonance.  For easier
    comparison, the values of $D$ are normalized to the
    resonance amplitude at zero detuning.
    \label{amplit_aS.fig}
}
\end{figure}

    These parameters are shown as functions of one-photon
detuning in different cells  in Figs.~\ref{angle_aS.fig} and
~\ref{amplit_aS.fig}.  We note that no deviation from the
symmetric Lorentzian form is observed for the EIT resonance in a
vacuum cell ($\phi=0$ for all detunings).  However, buffered cells
show the change from a symmetric transmission resonance (for
$\Delta=0$) to almost symmetric absorption resonance for
$\Delta\approx{700}$~MHz for $0.12$~Torr buffer gas pressure and
$\Delta\approx{1.4}$~GHz for $30$~Torr buffer gas pressure. After
reaching its maximum, the angle starts to decline again, although
we never observe the recovery of the symmetric EIT peak for larger
detunings.

\section{Theoretical analysis \label{theory.sec}}
\begin{figure}
 \includegraphics[angle=0, 
 	height=.3\textheight
	]
	 {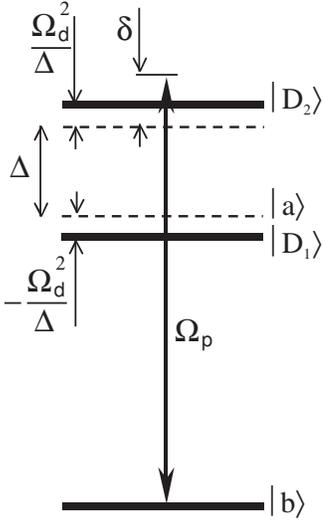}
 \caption{Dressed states picture for three level $\Lambda$ scheme. State
 $|D_1\rangle=\xi_1\left(|a\rangle + \frac{\Omega_d}{\Delta}|c\rangle\right) $
 has approximately the same energy as bare state $|a\rangle$, and state
 $|D_2\rangle=\xi_2\left(|c\rangle - \frac{\Omega_d^*}{\Delta}|a\rangle\right) $  is
 close to the two-photon resonance. The normalization coefficients are
$\xi_{1,2} \approx 1 + o(\frac{|\Omega_d|^2}{\Delta^2})$.
 EIT is observed for $\delta=0$ (maximum quantum interference), and absorption resonance corresponds to
 $\delta=-\frac{|\Omega_d|^{2}}{\Delta}$ (probe is resonant with $|D_2\rangle$).
 \label{dressedst.fig}     }
\end{figure}

Let us first give a brief pictorial explanation of the effect. The
observation of the narrow  absorption resonance  for  large laser
detunings may be qualitatively understood by using dressed-state
picture. It is known that the interaction of the drive field with
the transition  between levels  $|a\rangle$ and $|c\rangle$ can be
described as a splitting of  the excited state ~\cite{xiao95}.
Fig.~\ref{dressedst.fig} shows  the interaction scheme for  the
probe beam in the case  of the far-detuned drive field ($\Delta
\gg |\Omega_d|,  \gamma_{r}$).

For that conditions the transition between one of the dressed
states ($|D_1\rangle$ in our notation) and level $|b\rangle$
corresponds to a regular resonant absorption, and the transition
between another dressed state $|D_2\rangle$ and level $|b\rangle$
corresponds to a two-photon Raman transition. The destructive
interference of two dressed states occurs only in case of zero
two-photon detuning between probe and drive fields and corresponds
to the frequency with zero absorption (EIT). For the ideal case of
$\gamma_{bc} = 0$ (no ground state relaxation) and assuming that
all population is in level $|b\rangle$, the magnitudes of
absorption and Raman transition are the same while their widths are different:
the width of the former is determined by
the relaxation rate of optical coherence $\gamma$, while the width
of the Raman transition is given by $\gamma|\Omega_d|^2/\Delta^2$
and can be small for large $\Delta$. However, if spin relaxation
is not zero, the amplitude of resonance dramatically decreases
once the absorption resonance is narrower than this relaxation,
and for  $\gamma_{bc} \gg \gamma|\Omega_d|^2/\Delta^2$ the line
disappears. The addition of a buffer gas, therefore, improves the conditions 
for the observation of that narrow resonance: higher buffer gas pressure 
produces additional pressure broadening of the excited state ( i.e. makes 
$\gamma$ bigger) and at the same time restricts the diffusion of Rb atoms through 
the interaction region (i.e. makes $\gamma_{bc}$ smaller).
%
%
\footnote{Recently the sub-natural narrowing of the Doppler-free resonance
for far-detuned incoherent  fields in a vacuum cell has been demonstrated by U.\ D.\
Rapol \textit{et al.}~\cite{rapol03}}.

The dressed state picture provides a general idea about origin of the narrow
absorption resonances. Note, that the asymmetric lineshapes
observed in the experiment cannot be fitted by the combination of
two symmetric Lorentzian peaks
--- one positive due to coherent population trapping in the dark
state and one negative due to the enhanced absorption by the
bright state. The interference between two dressed state has to be taken
into account for proper description of the coherent effects ~\cite{lounis92}. 
\begin{figure} 
\includegraphics[angle=0,
    width=0.60\columnwidth]
    {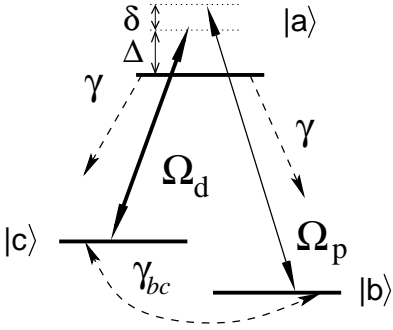}
    \caption{
    Three-level $\Lambda$ system.
    \label{levels-theory.fig}
}
\end{figure} 

    For more rigorous analysis we derive an expression for the absorption
coefficient of a weak probe field in Rb atoms. It has been shown
that a closed three-level $\Lambda$-scheme as shown in
Fig.~\ref{levels-theory.fig} provides satisfactory description of
coherent effects in alkali atoms~\cite{lee'03}. The propagation
equations for the density matrix elements are well-known (see,
e.g.\ \cite{lee'03}) and are the following:
\begin{eqnarray}
\label{rho_eit_eq_bia}
\dot{\rho}_{bb} & = &
    i \Omega_p^* \rho_{ab} - i \Omega_p \rho_{ba}
    + \gamma_r \rho_{aa} - \gamma_{bc} \rho_{bb} + \gamma_{bc} \rho_{cc}
        \\
\dot{\rho}_{cc} & = &
    i \Omega_d^* \rho_{ac} - i \Omega_{d} \rho_{ca}
    + \gamma_r \rho_{aa} - \gamma_{bc} \rho_{cc} + \gamma_{bc} \rho_{bb}
        \\
\dot{\rho}_{ab} &=& - \Gamma_{ab} \rho_{ab} + i \Omega_p (\rho_{bb} - \rho_{aa} )
        + i \Omega_d \rho_{cb}
        \\
\dot{\rho}_{ca}  &=& - \Gamma_{ca} \rho_{ca} + i \Omega_d^* (\rho_{aa} - \rho_{cc} )
        - i \Omega_p^* \rho_{cb}
        \\
\dot{\rho}_{cb} &=& - \Gamma_{cb} \rho_{cb} - i \Omega_p \rho_{ca}
        + i \Omega_d^* \rho_{ab}
    \label{rho_lambda_last_eq_bia}
\end{eqnarray}
where $\Omega_d=\wp_{ac}\mathrm{E_{p}}/\hbar$ and
$\Omega_p=\wp_{ab}\mathrm{E_{d}}/\hbar$ are the Rabi frequencies
of the drive and probe fields. The generalized decay rates are
defined as:
\begin{eqnarray}
\Gamma_{ba}&=&\gamma+i(\Delta+\delta)\,, \\
\Gamma_{ca}&=&\gamma+i\Delta\,,  \\
\Gamma_{bc}&=&\gamma_{bc}+i\delta\,.
\end{eqnarray}
Here $\gamma = \gamma_r+\gamma_{\mathrm{deph}}$ is the
polarization decay rate, $\gamma_r\simeq 2\pi\cdot 3$~MHz is the
radiative decay rate of the excited state,
$\gamma_{\mathrm{deph}}$ is the dephasing rate of the optical
transition due to non-radiative effects ($\gamma_{\mathrm{deph}}/p
\simeq 5$~MHz/Torr ~\cite{ottinger'75}); $\gamma_{bc}$ is the inverse
lifetime of the coherence between ground states $|b\rangle$ and
$|c\rangle$, which is determined by the diffusion time of Rb atoms
through the interaction region and nonhomogeneity of the magnetic
field due to imperfect screening. The presence of the buffer gas
affects values of both $\gamma_{bc}$ and $\gamma$. On one hand, it
allows for a longer ground-state coherence lifetime. On the other
hand it broadens the optical transition, since
$\gamma_{\mathrm{deph}}$ grows linearly with buffer gas
pressure~\cite{vanier_book}.

    Solving Eqs.(\ref{rho_eit_eq_bia})-(\ref{rho_lambda_last_eq_bia}) in a steady state regime and
assuming $|\Omega_p| \ll |\Omega_d|$, we obtain the following
expression for the linear susceptibility of the probe field:
\begin{equation}
\label{chi1}
\chi=i\kappa
\frac{\Gamma_{cb}(\rho^{(0)}_{aa}-\rho^{(0)}_{bb})+\displaystyle{\frac{|\Omega_d|^2}{\Gamma_{ca} }}(\rho^{(0)}_{aa}-\rho^{(0)}_{cc})}
{\Gamma_{ab}\Gamma_{cb}+|\Omega_d|^2}\,,
\end{equation}
Where $\kappa = \frac{3}{8\pi}N\lambda^2\gamma_r$, $N$ is the
$^{87}$Rb density and $\lambda$ is the wavelength of the probe field.

    The atomic population differences in the approximation
of strong driving field ($|\Omega_d|^2 \gg \gamma_{bc}\gamma$)
are:
\begin{eqnarray}
(\rho^{(0)}_{aa}-\rho^{(0)}_{bb}) &\simeq& -
\frac{\gamma_{bc}\Delta^2+
\gamma|\Omega_d|^2}{2\gamma_{bc}\Delta^2+
\gamma|\Omega_d|^2} \\
(\rho^{(0)}_{aa}-\rho^{(0)}_{cc}) &\simeq& -
\frac{\gamma_{bc}(\Delta^2+\gamma^2)}{2\gamma_{bc}\Delta^2+
\gamma|\Omega_d|^2}\,.
\end{eqnarray}
By substituting these expressions into Eq.~(\ref{chi1}), we find
the absorption coefficient $\alpha=Im\{\kappa\}$ as a function of two-photon
detuning $\delta$:
\begin{equation} \label{abs1}
\alpha=\frac{\kappa}{\gamma^2+\Delta^2}\frac{\gamma_{bc}\Delta^2+
\gamma|\Omega_d|^2}{2\gamma_{bc}\Delta^2+
\gamma|\Omega_d|^2}\frac{\gamma_{bc}|\Omega_d|^2+
\gamma\delta^2}{\tilde \gamma^2+(\delta-\delta_0)^2}\,,
\end{equation}
where
\begin{equation} \label{delta0}
\delta_0=|\Omega_d|^2\frac{\Delta}{\gamma^2+\Delta^2}
\end{equation}
is the ac-Stark shift of the excited state, and
\begin{equation} \label{width-theory}
\tilde \gamma = \frac{\sqrt{\gamma^2|\Omega_d|^4 +
\gamma_{bc}^2\Delta^2(\gamma^2+\Delta^2)}}{\gamma^2+\Delta^2}
\end{equation}
is the effective width of the two-photon transmission resonance.

    Using Eq.~(\ref{abs1}) for the absorption coefficient we
can now find expressions for the coefficients $A$, $B$, and $C$
in Eq.~(\ref{res_shape_fit}) which describe the line-shape
of the two-photon resonance for the probe field propagating
through a medium of length $L$.  For the moment we restrict
ourselves to the case of optically thin media, so that $I_{p}(z) =
I_p(0)\cdot e^{-\alpha z} \simeq I_p(0)(1-\alpha z)$:
\begin{eqnarray}
A &=& \kappa L \eta
\frac{|\Omega_d|^2}{\gamma^2+\Delta^2}\frac{\gamma|\Omega_d|^2(\gamma^2-\Delta^2)
- \gamma_{bc}(\gamma^2+\Delta^2)^2}{\gamma^2|\Omega_d|^4 +
\gamma_{bc}^2\Delta^2(\gamma^2+\Delta^2)} \label{a}\nonumber\\
&~&\\
B &=& - \kappa L \eta \frac{\Delta}{\gamma^2+\Delta^2} \label{b} \\
C &=& 1-\kappa L \eta \ \frac{\gamma}{\gamma^2+\Delta^2} \label{c} \\
\eta &=& \frac{\gamma_{bc}\Delta^2+
\gamma|\Omega_d|^2}{2\gamma_{bc}\Delta^2+ \gamma|\Omega_d|^2} \,.
\end{eqnarray}

    It is easy to see that the coefficient $C$
is well approximated by the absorption of a weak probe field in a
two-level scheme. The only difference is the coefficient $\eta$
which describes the redistribution of the atomic population
between ground levels due to optical pumping:  in the case of a
small one-photon detuning ($\Delta \ll
|\Omega_d|\sqrt{\gamma/\gamma_{bc}}$) almost all atoms are in the
state $|b\rangle$ and therefore $\eta = 1$, and if the laser is
far-detuned, the populations of the levels $|b\rangle$ and
$|c\rangle$ are almost the same and $\eta=1/2$.

Coefficient $A$, which describes the symmetric component of the
resonant lineshape is a symmetric function of the one-photon
detuning. Its sign changes as $\Delta$ gets larger: $A$ is
positive for small detunings, then it becomes negative at $\Delta
\approx \gamma - 2\gamma_{bc}\gamma^2/|\Omega_d|^2$ (i.e. the
resonance becomes absorptive). Coefficient $B$ is an odd function
of $\Delta$, and exactly zero for $\Delta=0$.
%
%
\begin{figure} 
\includegraphics[angle=0,
    width=1.00\columnwidth]
{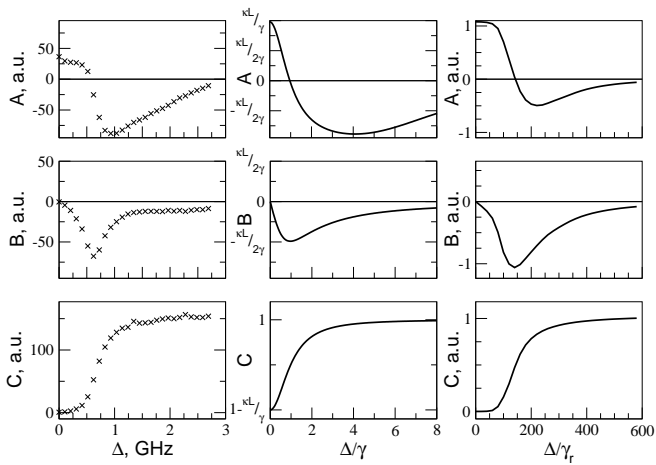}
\caption{
    Coefficients $A$, $B$, and $C$ describing the line-shape
    of the coherent resonance Eq.~(\ref{res_shape_fit}):
    extracted from the experimental data for $^{87}$Rb cell
    with $30$~Torr (left column), calculated using
    Eqs.~(\ref{a}--\ref{c}) (middle column),
    and obtained by numerical modelling (right column).
    \label{abc-comp.fig}
}
\end{figure} 

    A comparison of coefficients $A$, $B$, and $C$ given
by Eqs.~(\ref{a}--\ref{c}), with those obtained by fitting our
experimental data for the Rb cell with $30$~Torr of Ne is
presented in Fig.~\ref{abc-comp.fig}.  One can immediately see
that the theoretical formulae qualitatively describe the
dependence of the coefficients as functions of one-photon
detuning, although they are not accurate enough for a quantitative
analysis.  There are several reasons for this.  On one hand, we
will show below that the thermal velocity distribution of Rb atoms
is very important and has to be taken into account.
On the other hand, we assumed for simplicity that the medium is optically
thin which is not the case.  One of the consequences of using optically
thick medium is density narrowing of the EIT resonances
which will be discussed in the following Section.

We perform a numerical simulation of the interaction of the drive
and probe field with the three-level $\Lambda$ system considered
above. Our model takes into account attenuation of both drive and
probe fields as they propagate through Rb vapor. The thermal
motion of atoms also implies that atoms with different velocities
``see'' the electromagnetic fields at shifted frequencies, and the
final susceptibility has to be averaged over the Maxwell velocity
distribution:
\begin{equation} \label{maxwell_ave}
\chi(\Delta)=\frac{1}{\sqrt{\pi}ku}
\int_{-\infty}^{+\infty}{\chi(\Delta-kv)e^{-\frac{(kv)^2}{(ku)^2}}
\mathrm{d}(kv)}\,,
\end{equation}
where $k=\nu_{d}/c$ is the drive field wave-vector, and $u =
\sqrt{2k_BT/M}$ is the most probable thermal speed (here $T$ is
the temperature of the vapor, $M$ is the mass of Rb atom and $k_B$
is the Boltzmann constant). The residual Doppler broadening of the
ground-state transition ($\Delta ku$) may be neglected due to
Lamb-Dicke effect ~\cite{helm'01}.

\begin{figure*} 
\includegraphics[angle=0,
    width=1.20\columnwidth]
{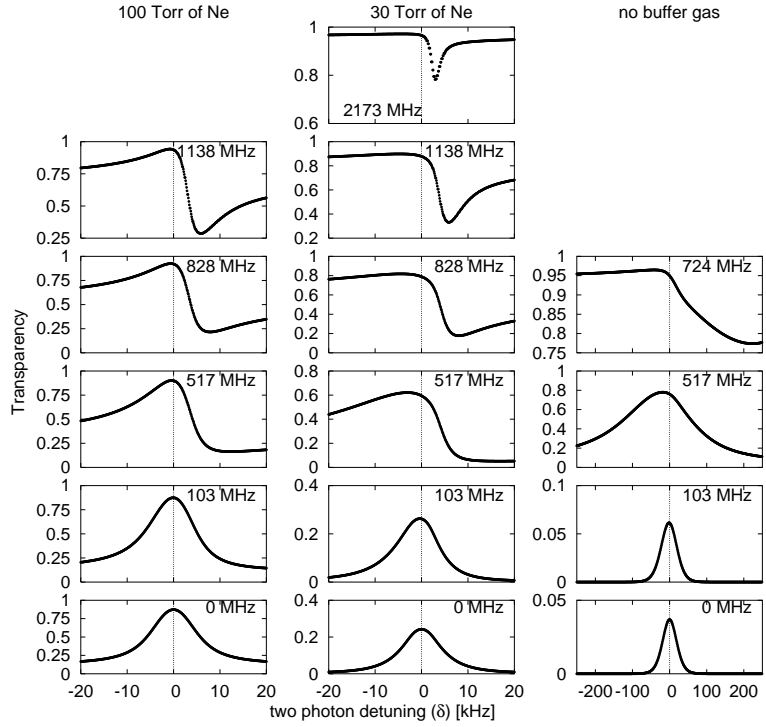}
\caption{
    Calculated probe field transmission spectra for the conditions
    corresponding the experimental data in Fig.~\ref{eit_shape_example}.
    We use the following parameters: $\gamma_{\mathrm{deph}} = 450
    $~MHz, $\gamma_{bc}=0.5$~kHz for the cell with $100$~Torr of Ne;
    $\gamma_{\mathrm{deph}} = 150$~MHz, $\gamma_{bc}=0.7
    $~kHz for the cell with $30$~Torr of Ne;
    and $\gamma_{\mathrm{deph}} = 0$\,
    $\gamma_{bc}=30$~kHz for the cell without a buffer gas.
    These values are
    in good agreement with published results for collisional
    dephasing~\cite{ottinger'75,happer72}. For all graphs $ku=250$~MHz,
    $\Omega_d=2.5$~MHz,
    and $\Omega_p=0.5$~MHz.  These values are reasonable close to estimated
actual Rabi frequencies
    $\Omega_d=2.1$~MHz  and $\Omega_p=0.7$~MHz. The small deviation between
them can be attributed to the nonuniform spatial distribution
    of the laser intensities.
    \label{eit_shape_calc}
}
\end{figure*}

We can reproduce experimental spectra very accurately (compare the
experimental spectra presented in Fig.~\ref{eit_shape_example}
with the the results of numerical simulations in
Fig.~\ref{eit_shape_calc}) except for the case of cells with small
amount of a buffer gas ($<1$~Torr). This limit corresponds to the
regime where the mean free path of Rb atoms is comparable with the
size of the interaction region. Under these conditions the motion
of Rb atoms in and out of the laser beam has to be taken into
account when calculating the ground state coherence lifetime.
Unfortunately, this mechanism is beyond the theoretical model
used in our numeric simulations. For that reason we do not show
any calculated transmission spectra for the cell with $0.12$~Torr
of Kr. The results of the numerical simulation for the parameters $A$,
$B$ and $C$ are shown in Fig.~\ref{abc-comp.fig}c.

%

\begin{figure}   \center
  \includegraphics[
  width=1.00\columnwidth
 ]
 {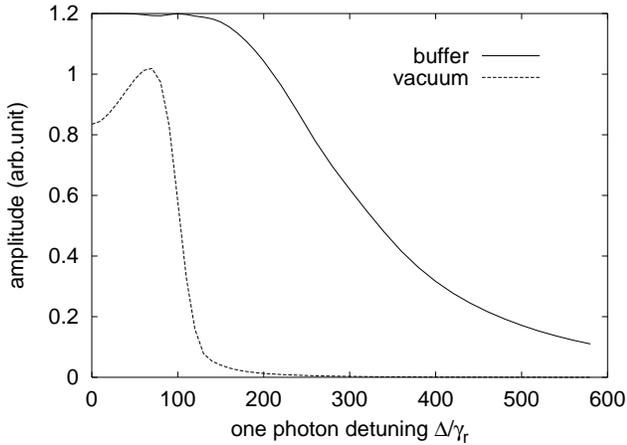}
 \caption{
    Numerical calculation of the resonance strength ($D$) vs
    one-photon detuning for a probe field propagating in a
    medium with Ne buffer gas and with vacuum.  One-photon detuning
    and resonance width are given in units of $\gamma_r$\,.
 \label{buf_vac_signal_theor}
 }
\end{figure}

\begin{figure}
 \center
 \includegraphics[
  width=1.00\columnwidth
 ]
 {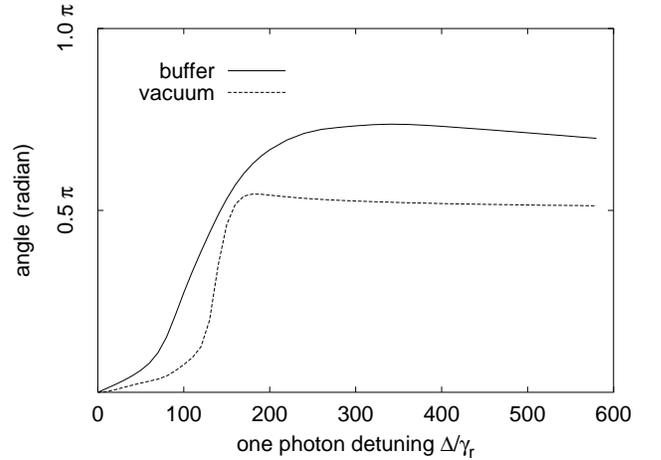}
\caption{
    Numerical calculation of angle ($\phi$) vs one-photon
    detuning for a probe field propagating in a medium
    with Ne buffer gas and with vacuum.  One-photon detuning and
    resonance width are given in units of $\gamma_r$\,.
 \label{buf_vac_angle_theor}
}
\end{figure}

 The calculated values of the
resonance amplitude $D$ and the ratio $\phi$ are shown in
Figs.~\ref{buf_vac_signal_theor} and \ref{buf_vac_angle_theor}.
These demonstrate that the inversion of the EIT resonance occurs
in the cell with a buffer gas, whereas  no absorption resonances
ever appear in the cell without buffer gas. Although a noticeable
asymmetry of the resonance is expected for large laser detuning
($\Delta > 100\gamma_r$)\,,
the amplitude of the resonances is very small and hardly detectable. %

\begin{figure} 
\includegraphics[angle=0,
    width=1.00\columnwidth]
    {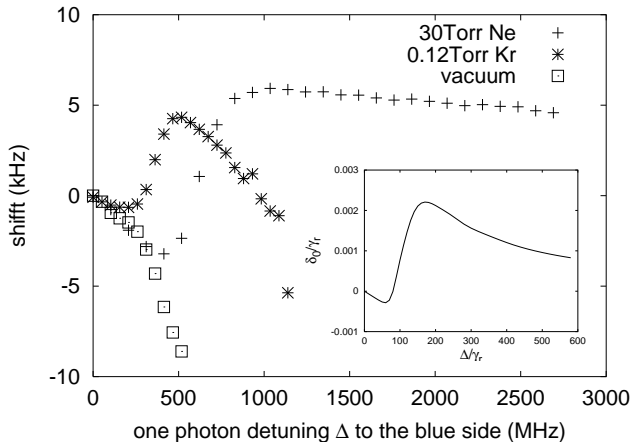}
    \caption{
    Two photon resonance shift $(\delta_0)$ as a function
    of one-photon detuning for the vacuum cell (squares),
    cell with $0.12$~Torr of Kr (stars) and with $30$~Torr
    of Ne (crosses).  \textit{Inset:} the result of the
    numerical simulation for the $30$~Torr cell.
    \label{res_shift.fig}
}
\end{figure}

    It is also easy to
see in Fig.~\ref{eit_shape_example} that for non-zero detuning
$\Delta$ the centers of both EIT and buffer-gas induced absorption
resonances are shifted from zero two-photon detuning. One of the
reasons for this effect is light shifts of the atomic levels, as
shown by Eq.~(\ref{delta0}).  However, this expression fails to
describe the behavior of the resonance shift measured
experimentally (Fig.~\ref{res_shift.fig}). A more realistic
resonance shift as a function of laser detuning is obtained by
numerical simulation if Doppler averaging is performed
(Fig.~\ref{res_shift.fig}, inset).

    We note that both the prediction of the theoretical
model and the result of the numerical simulations provide only
qualitative agreement with the experimental results. There are
several major effects which are not considered here. For example,
neither hyperfine structure of the excited state nor Zeeman
substructure of all states are taken into consideration, although
this may have a profound effect on the coherent interaction.  In
addition, no influence of the four-wave mixing processes is taken into
account.

For the better understanding of the conditions for which the enhanced
absorption resonance occurs, we numerically compare the resonance lineshape at
a large one-photon detuning ($\Delta=250\gamma_r$) for the cases of a cell with
and without a buffer gas. We see that the
absorption resonances appear only for large enough drive laser
power $\Omega_d$ in the cell with a buffer gas. In the cell without any
buffer gas the resonance remains dispersion-like ($\phi \approx
\pi/2$) and vanishingly weak.


\section{Influence of the buffer gas on four-wave mixing}

So far we have completely ignored the presence of a Stokes field
$\mathrm{E_S}$ in the medium. Many previous
publications~\cite{lukin'97prl, harris99prl, kash99,
fleischhauer'02} have shown that dense coherent media contribute
to significant enhancement of nonlinear processes.  In our
experiment the strong drive field applied to the $F=1$ level is
scattered by the ground-state coherence, which leads to the
appearance of narrow resonances in the Stokes field transmission
with the width determined by the ground-state coherence relaxation
rate. For near resonant probe and drive fields ($\Delta \approx
0$)the $\Lambda$ system formed by the Stokes and the drive field
applied to $F=1$ ground state level is far detuned (see
Fig.~\ref{levels.fig}a), and the effects of the associated
ground-state coherence on the probe field are negligible. However,
all four fields have to be taken into account in the case of large
laser detuning, comparable with the ground state hyperfine
splitting.
\begin{figure} 
\includegraphics[angle=0,
    width=1.00\columnwidth]
    {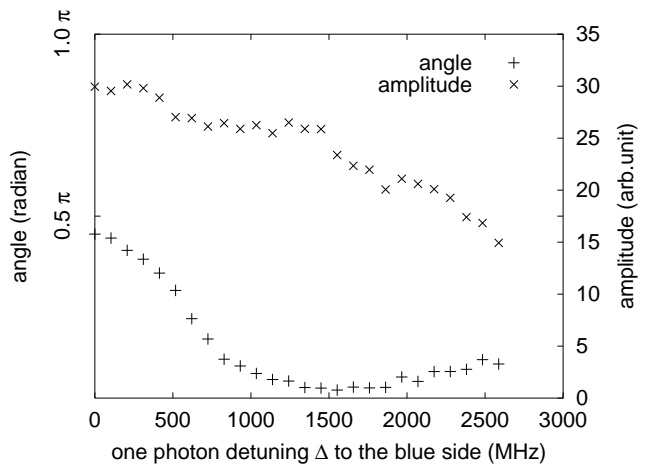}
\caption{
    Angle ($\phi$) and amplitude ($D$) of the two photon
    resonance for the generated Stokes field.
    \label{angle_and_amplit_S}
}
\end{figure} 

The amplitude and the angle $\phi$ for the Stokes field transmission
resonance in the cell with $30$~Torr of Ne are shown in
Fig.~\ref{angle_and_amplit_S}. Please note that for $\Delta=0$ we
observe a dispersion-like lineshape ($\phi\approx 0.5\pi$), which
is different from the previously published experimental results
for the Stokes field generation initiated by spontaneous
photons~\cite{mikhailov2002}. As the laser detuning $\Delta$
increases the lineshape of the Stokes field transmission changes, and
the resonance is transformed into a symmetric transmission peak
($\phi\approx 0$).

\section{Width of the probe and Stokes resonances}

\begin{figure} 
\includegraphics[angle=0,
    width=1.00\columnwidth]
    {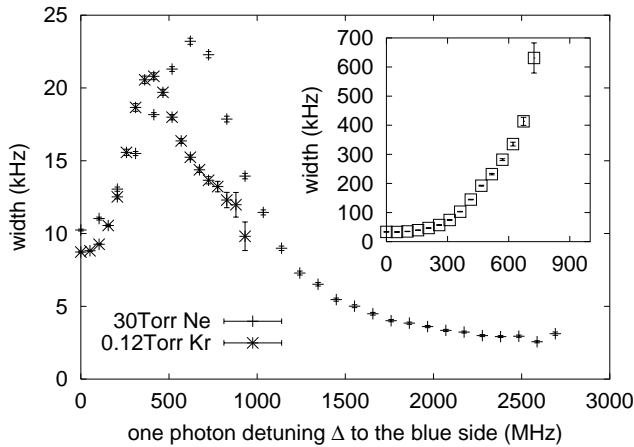}
    \caption{
    Width of the two photon resonance $\tilde\gamma$
    as a function of one-photon detuning $\Delta$ for
    $^{87}$Rb cell with $30$~Torr of Ne (cross), $0.12$~Torr
    of Kr (x), and without buffer gas (squares, inset).
    The minimum width of the EIT resonance in the vacuum
    cell with no buffer gas is $\tilde\gamma (\Delta=0)
    = 17~\mathrm{kHz}$.
\label{res_width}
}
\end{figure}

   The measured widths of the two-photon resonance $\tilde
\gamma$ as functions of the one-photon detuning $\Delta$ are shown
in Fig.~\ref{res_width} for cells with different buffer gas
pressure.  Again we see the difference between cells with and
without a buffer gas:  in the latter case the width of the EIT
resonance increases significantly with one-photon detuning.
However, for the buffered cells the resonance is broadened only
near atomic resonance, but for larger detuning its width actually
decreases with $\Delta$.

\begin{figure}
    \center
    \includegraphics[
        width=1.00\columnwidth
    ]
    {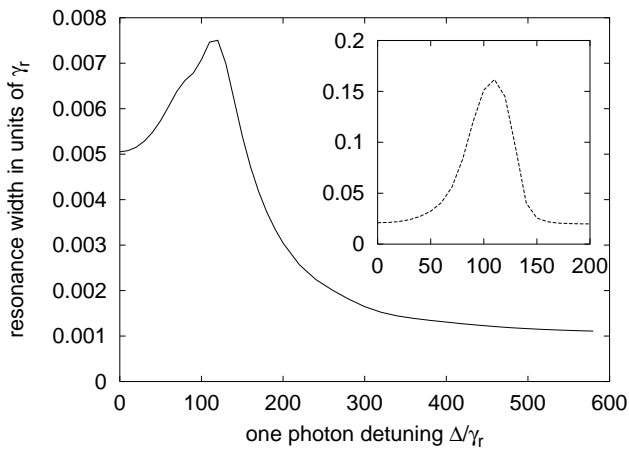}
    \caption{
    Numerical calculation of resonance width
    ($2\tilde{\gamma}$) vs one-photon detuning for a probe
    field propagating in a medium with Ne buffer gas and with vacuum
    (inset).
        \label{buf_vac_width_theor}
    }
\end{figure}

Fig.~\ref{buf_vac_width_theor} demonstrate that we may be able observe the 
narrowing of resonances for large detuning even
in a vacuum cell.   However according to Fig.~\ref{buf_vac_signal_theor}
at the point when narrowing takes place ($\Delta/\gamma_r >100$) the amplitude
of the resonance is so small that it is extremely hard to detect in the
experiment.

 Narrow resonances with good signal-to-noise ratio
are important for many applications.  For example, narrow EIT
resonances are used for precision metrology~\cite{wynands'98,
wynands'02, budker'00} and atomic clocks~\cite{hollberg'02,
merimaa'03}.  Our experiments demonstrate that the coherent
absorption resonances, observed for the far-detuned $\Lambda$
system, may have more attractive characteristics in terms of the
resonance width and amplitude than the EIT resonances observed for
the zero detuning.  For example, in the cell with $30$~Torr of Ne
the amplitude of the absorption resonance for $\Delta =
1.2$--$2$~GHz is larger than that of the EIT resonance, while its
width is narrower (for example, $\tilde\gamma(0)/\tilde
\gamma(2$GHz$)\approx 3.3$).

%

According to Eq.~(\ref{width-theory}) in the strong laser field
limit ($|\Omega_d|^2 \gg \gamma_{bc}\gamma$) the width of the EIT
resonance for small $\Delta$ does not depend on the ground-state
coherence decay rate and is determined by power broadening:
$\tilde \gamma \approx |\Omega_d|^2/\gamma$, as in previous
studies~\cite{javan'02}. The resonance width decreases with
one-photon detuning, and for $\Delta \gg \gamma$ it drops as
$1/\Delta^2$. Ultimately for $\Delta \gg |\Omega_d|
\sqrt{\gamma/\gamma_{bc}}$\,, the width of the resonance is
determined by the coherence decay rate $\gamma_{bc}$.

\begin{figure} 
\includegraphics[angle=0,
    width=1.00\columnwidth]
    {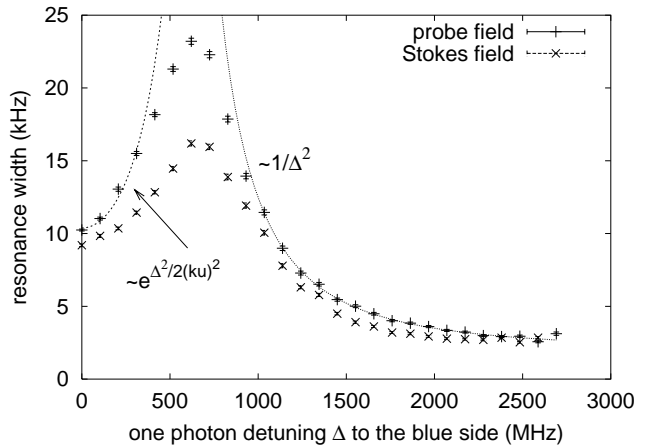}
    \caption{
    Width of the two photon resonance $(\gamma)$ for the probe
    and generated Stokes field.
    \label{res_width_both}
}
\end{figure}

To describe the resonance width more carefully we again have
to take into account the velocity distribution of Rb atoms in the
cell. Also the variation of the laser fields intensities as
they propagate through the optically dense Rb vapor becomes
important; for example, the resonance width is reduced due to the density
narrowing~\cite{lukin97prl,sautenkov99las}. We note that both
effects are not very important for the far-detuned optical
fields ($\Delta \gg \gamma,ku$) since the absorption of the medium
is small, and the parameters describing the resonance lineshape do
not change noticeably within the Doppler contour. Thus, for large
detuning the dependence of the resonance width predicted by
Eq.~(\ref{width-theory}) is in good agreement with the
experimental points, as shown in Fig.~\ref{res_width_both}.

To describe how the resonance width changes for small $\Delta$ we
should take both propagation and Doppler averaging into account.
Since it is virtually impossible to do that exactly, we make a few
simplifications that allow us to roughly predict the dependence of
the resonance width on one-photon detuning. First, we assume that
the main contribution to the observed resonance comes from the
velocity subgroup for which $\Delta-kv \simeq 0$, i.e. the
coupling fields are resonant with the atomic transitions. Then the
main consequence of the laser detuning is the reduction of the
number of atoms in this velocity group:
\begin{equation} \label{density}
N(\Delta)\propto Ne^{-\frac{\Delta^2}{(ku)^2}}.
\end{equation}
Under these conditions the output intensity $I_{\mathrm{out}}$ is
given by
\begin{equation}
I_{\mathrm{out}} \simeq I_{\mathrm{in}} e^{-\kappa
L\frac{\gamma_{bc}}{|\Omega_d|^2}}e^{-\kappa
L\frac{\gamma\delta^2}{|\Omega_d|^4}} \,.
\end{equation}
In this expression the first term represents the residual
absorption under EIT conditions, and the second one describes
the shape of the peak as a function of two-photon detuning.
It is easy to see that the width of the resonance in this case
is inversely proportional to the atomic density \cite{lukin97prl}:
\begin{equation}
\tilde \gamma_D(N) \simeq
\frac{|\Omega_d|^2}{\sqrt{\gamma\gamma_r}}\left
(\frac{3}{8\pi}N\lambda^2L\right )^{-1/2} \,.
\end{equation}
Substituting the density of the resonant atoms given by
Eq.~(\ref{density}) we obtain the resonance width:
\begin{equation} \label{width_near_res}
\tilde \gamma_D(\Delta) \propto
\frac{|\Omega_d|^2}{\sqrt{\gamma\gamma_r}}\frac{1}{\sqrt{NL}}
\, e^{\frac{\Delta^2}{2(ku)^2}}.
\end{equation}
Fig.~\ref{res_width_both} shows that this formula works only for a
small frequency region in the vicinity of one-photon resonance
where the resonance is nearly symmetric and deviates for larger
$\Delta$, which is consistent with the assumptions we made.

    The width of the Stokes field resonance is also shown
in Fig.~\ref{res_width_both}.  It has approximately the
same shape as the curve for the probe field, but it is worth
mentioning that its width can be narrower than that of the
probe field while having a similar resonance amplitude.

\section{Enhanced absorption due to Zeeman coherence}

    Let us now consider another kind of coherence effect
originating from coherent population trapping, namely nonlinear
magneto-optical polarization rotation.  An extensive review of
this effect is given in Ref.~\cite{budker2002rmp}.  In this case,
the long-lived coherence is created among ground-state Zeeman
sublevels by a single electromagnetic field (in general
elliptically polarized but usually taken as linear in most
experiments and analyses) where the $\Lambda$ link(s) are formed
by the two opposite circularly polarized components $E_\pm$ of the
input field (Fig.~\ref{levels.fig}b).  In this configuration the
transmission is very sensitive to the applied magnetic field,
which shifts sublevels with different magnetic quantum number $m$.
An applied magnetic field $B$ therefore creates a two-photon
detuning $\delta$ proportional to the magnetic field:
$\delta=2\mu_BB/\hbar$, where $\mu_B$ is the Bohr magneton. In
particular, the polarization of the incident laser field is
changed dramatically as a result of the steep dispersion
associated with CPT~\cite{sautenkov'00, novikova'01ol}.  This
effect, known as the nonlinear Faraday effect, or nonlinear
magneto-optical polarization rotation, has been extensively
studied over the last decades in atomic beams~\cite{theobald'89,
schuh'93} and in glass cells with or without anti-relaxation
coating~\cite{drake'88, barkov'89, baird'89, chen'90a,
kanorsky'95, budker'98, budker'99ajp, sautenkov'00,
novikova'01ol}.

    There are a number of publications which demonstrate
strong influence of a buffer gas on the amplitude and lineshape of
polarization rotation and EIT resonances~\cite{zibrov'01ol,
novikova'02apl}.  Here we show that presence of a
buffer gas also allows observation of narrow enhanced absorption
peaks for buffer gas pressure higher than $3$~Torr.

\begin{figure}
\includegraphics[angle=0,width=1.00\columnwidth]{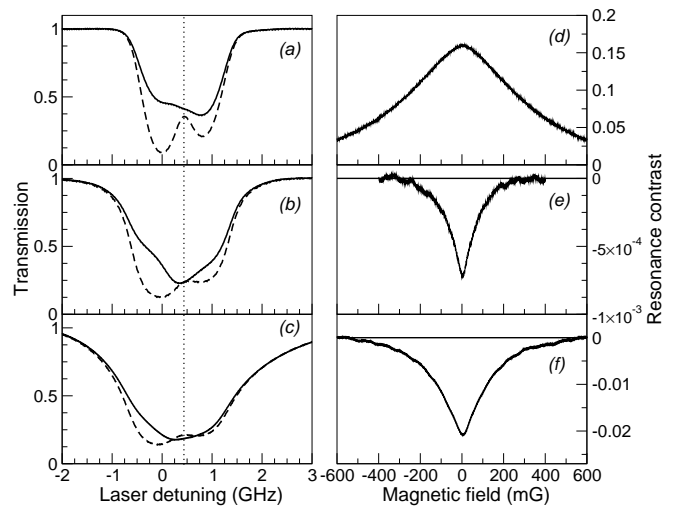}
\caption{
    \textit{Left column}: transmission through the $^{87}$Rb
    cells with (a) $1$~Torr; (b)$3$~Torr (c) $10$~Torr of
    Ne buffer gas as the laser frequency is swept across
    the $F=2 \rightarrow F'$ transitions.  Solid line:
    transmission at zero magnetic field.  In this case the
    ground state coherence is unperturbed and we denote
    the transmission as $\mathrm{T_{\mathrm{coh}}}$.
    Dashed line: transmission at large magnetic field
    ($\delta \gg \gamma_0$).  In this case the coherence
    is destroyed, and we denote the transmission as
    $\mathrm{T_{\mathrm{lin}}}$.  \textit{Right column}:
    The peak contrast in transmission as a function of
    magnetic field (two-photon detuning) for fixed laser
    frequency (shown as a dotted line on the previous
    graphs) for the same cells.  The normalization of
    the output signal is $C=(\mathrm{T_{\mathrm{coh}}} -
    \mathrm{T_{\mathrm{lin}}})/\mathrm{T_{\mathrm{coh}}}$.
    Thus $C>0$ is a manifestation of EIT shown in part (d),
    and $C<0$ indicates enhanced absorption as seen in (e)
    and (f).  Because of the difference in geometrical
    sizes of the cells, the atomic density is adjusted
    for each case so that $\mathrm{T_{\mathrm{lin}}}
    \sim 20\%$.  Laser power is $2$~mW.
    \label{zeeman.fig}
}
\end{figure}

    The transmission spectrum of $^{87}$Rb $5S_{1/2}F=2
\rightarrow 5P_{1/2}~F'=1,2$ consists of two transitions,
partially resolved within the Doppler contour as shown in
Fig.~~\ref{levels.fig}b. Zeeman coherence can be created on both
transitions so that linear absorption is suppressed for a linearly
polarized electromagnetic field even if the laser frequency
differs somewhat from the exact atomic transition frequency.  For
example, if the laser is tuned exactly between the two transitions
($\Delta=406$~MHz), the transmission is still enhanced by $40\%$.

    This is true, however, only if there is no buffer gas
present in the cell.  The transmission spectra for different
pressures of Ne buffer gas are shown in Fig.~\ref{zeeman.fig}a, b,
and c.  One can easily see that the effect of EIT deteriorates at
higher buffer gas pressures.  Moreover, the enhanced absorption
appears for the frequency region between the two transitions. When
magnetic field is varied the width of the observed resonances is
similar to that of EIT peak as shown in Fig.~\ref{zeeman.fig}d, e,
and f, which indicates that this effect is due to the ground-state
coherence. Similar resonances have been observed in the cells with
anti-relaxation wall coating \footnote{D.\ Budker, private
communication.}

    Here we should emphasize that the reason for the
enhanced absorption resonances described in this Section is
quite different from the effects described in the previous
sections, despite of their similarities (both effects
are observed only in buffered Rb cells and for nonzero
one-photon detunings).  In the case of Zeeman coherence,
the nonlinear enhanced absorption may be explained by the
interplay of the matrix elements of the transitions involved
in the $\Lambda$ links.  As the laser frequency changes,
dark states can be created for both $F=2 \rightarrow F'=1$
and  $F=2 \rightarrow F'=2$  transitions.  However, because
of the difference in relative sign in the transition matrix
elements~\cite{vanier_book}, these dark states are orthogonal:
\begin{eqnarray}
|D\rangle_{2\rightarrow 1} &=& \frac{1}{\sqrt{2}}\left(|m=1\rangle+|m=-1\rangle \right); \label{dark21} \\
|D\rangle_{2\rightarrow 2} &=&
\frac{1}{\sqrt{2}}\left(|m=1\rangle-|m=-1\rangle \right);
\label{dark22}
\end{eqnarray}
Strictly speaking, because of this difference perfect CPT is not
possible in this double-$\Lambda$ system.  However, the natural
width of both transitions ($\gamma_r \simeq 3$~MHz) is much
smaller than their hyperfine splitting, and for the laser resonant
with one transition, the disturbance introduced by the other
transition may be neglected.  For the cell without buffer gas or
anti-relaxation coating, atoms do not change their velocity while
moving through the laser beam and interact with the same atomic
transition. Therefore we observe two Gaussian-shaped transmission
peaks as the laser frequency is swept across the $F=2 \rightarrow
F'$ transitions.

In the presence of a buffer gas the situation is different.  Since
the hyperfine structure of the excited level is not completely
resolved under Doppler broadening, there is a finite probability
that an atom, optically pumped into the dark state on one
transition, may later change its velocity and become resonant with
the other transition. Because of the sign difference, the
previously prepared dark state becomes a bright state for that
transition, and an atom absorbs light more readily than the one
without coherence.  Thus, for the symmetric scheme as shown in
Fig.~\ref{levels.fig}b, we would expect complete destruction of
coherence for the laser tuned exactly between two atomic
transitions, so that the probabilities of an atom interacting with
either of them are equal. In reality, however, the $F=2
\rightarrow F'=1$ transition is stronger then the $F=2 \rightarrow
F'=2$ transition, and this imbalance results in the enhanced
absorption for a small region of laser frequencies.

\section{Summary}

    We have demonstrated significant changes of the EIT resonance
in warm Rb vapor with a buffer gas.  We showed that the resonance
lineshape changes from symmetric transparency peak to a
dispersion-like signal and then to nearly symmetric absorption
peak as one-photon detuning of the laser fields from atomic
transition increases. We also found that for large enough
detuning, the resonance width was limited by the ground level
coherence decay rate ($\gamma_{bc}$), and there is a region of
laser frequencies with higher resonance contranst than for zero
one-photon detuning, while the width of the resonance is narrower.
Observation of this absorption-like resonance is power dependent
and takes place only for high enough drive power.  We have also
developed a simple theory and performed the numerical stimulations
that qualitatively describe the observations, and we have
accounted for the origin of the discrepancies between those and
the experimental data. We compared the behavior of the probe and
Stokes field for large one-photon detuning, and showed that the
resonance width of these two fields behaves similarly, while the
shapes of the resonances are different.

    We have also shown that an induced absorption resonance
appears with one-photon detuning in the Hanle configuration, although
its origin differes from the bichromatic case. In the
Hanle configuration, the enhanced absorption results from the
destructive interference between the dark states created via
different excited states in the presence of velocity changing
collisions.

    The effects reported here differ from previously
observed Electromagnetically Induced Absorption resonances and
the absorption Hanle effect.  Both of those cases have strict
requirements on selection rules that are not required or met
for our observations.


\section{Acknowledgments}

    We thank V.\ A.\ Sautenkov, D. Budker, A.\ B.\ Matsko, A.\ Zhang,
and M.\ O.\ Scully for the useful and stimulating discussions, and M.\ Klein
for careful reading of the manuscript. Financial support is provided by 
Office of Naval Research.




\begin{thebibliography}{73}
\expandafter\ifx\csname natexlab\endcsname\relax\def\natexlab#1{#1}\fi
\expandafter\ifx\csname bibnamefont\endcsname\relax
  \def\bibnamefont#1{#1}\fi
\expandafter\ifx\csname bibfnamefont\endcsname\relax
  \def\bibfnamefont#1{#1}\fi
\expandafter\ifx\csname citenamefont\endcsname\relax
  \def\citenamefont#1{#1}\fi
\expandafter\ifx\csname url\endcsname\relax
  \def\url#1{\texttt{#1}}\fi
\expandafter\ifx\csname urlprefix\endcsname\relax\def\urlprefix{URL }\fi
\providecommand{\bibinfo}[2]{#2}
\providecommand{\eprint}[2][]{\url{#2}}

\bibitem[{\citenamefont{Arimondo and Orriols}(1976)}]{arimondo'76}
\bibinfo{author}{\bibfnamefont{E.}~\bibnamefont{Arimondo}} \bibnamefont{and}
  \bibinfo{author}{\bibfnamefont{G.}~\bibnamefont{Orriols}},
  \bibinfo{journal}{Nuovo Cimento Lett.} \textbf{\bibinfo{volume}{17}},
  \bibinfo{pages}{333 } (\bibinfo{year}{1976}).

\bibitem[{\citenamefont{Agapyev et~al.}(1993)\citenamefont{Agapyev, Gornyi,
  Matisov, and V.Rozhdestvenskii}}]{agapyev'93}
\bibinfo{author}{\bibfnamefont{B.~D.} \bibnamefont{Agapyev}},
  \bibinfo{author}{\bibfnamefont{M.~B.} \bibnamefont{Gornyi}},
  \bibinfo{author}{\bibfnamefont{B.~G.} \bibnamefont{Matisov}},
  \bibnamefont{and}
  \bibinfo{author}{\bibfnamefont{Y.}~\bibnamefont{V.Rozhdestvenskii}},
  \bibinfo{journal}{Usp.\ Fiz.\ Nauk} \textbf{\bibinfo{volume}{163}},
  \bibinfo{pages}{1} (\bibinfo{year}{1993}).

\bibitem[{\citenamefont{Arimondo}(1996{\natexlab{a}})}]{arimondo'96po}
\bibinfo{author}{\bibfnamefont{E.}~\bibnamefont{Arimondo}},
  \bibinfo{journal}{Progress in Optics} \textbf{\bibinfo{volume}{XXXV}},
  \bibinfo{pages}{259} (\bibinfo{year}{1996}{\natexlab{a}}).

\bibitem[{\citenamefont{Scully and Zubairy}(1997)}]{scullybook}
\bibinfo{author}{\bibfnamefont{M.~O.} \bibnamefont{Scully}} \bibnamefont{and}
  \bibinfo{author}{\bibfnamefont{M.~S.} \bibnamefont{Zubairy}},
  \emph{\bibinfo{title}{Quantum Optics}} (\bibinfo{publisher}{Cambridge
  University Press, Cambridge, UK}, \bibinfo{year}{1997}).

\bibitem[{\citenamefont{Harris}(1997)}]{harris'97pt}
\bibinfo{author}{\bibfnamefont{S.~E.} \bibnamefont{Harris}},
  \bibinfo{journal}{Phys.\ Today} \textbf{\bibinfo{volume}{50}},
  \bibinfo{pages}{36} (\bibinfo{year}{1997}).

\bibitem[{\citenamefont{Marangos}(1998)}]{marangos'98}
\bibinfo{author}{\bibfnamefont{J.~P.} \bibnamefont{Marangos}},
  \bibinfo{journal}{J.\ Mod.\ Opt.} \textbf{\bibinfo{volume}{45}},
  \bibinfo{pages}{471 } (\bibinfo{year}{1998}).

\bibitem[{\citenamefont{Harris et~al.}(1992)\citenamefont{Harris, Field, and
  Kasapi}}]{harris'92}
\bibinfo{author}{\bibfnamefont{S.~E.} \bibnamefont{Harris}},
  \bibinfo{author}{\bibfnamefont{J.~E.} \bibnamefont{Field}}, \bibnamefont{and}
  \bibinfo{author}{\bibfnamefont{A.}~\bibnamefont{Kasapi}},
  \bibinfo{journal}{Phys.\ Rev.\ A} \textbf{\bibinfo{volume}{46}},
  \bibinfo{pages}{R29 } (\bibinfo{year}{1992}).

\bibitem[{\citenamefont{Schmidt et~al.}(1996)\citenamefont{Schmidt, Wynands,
  Hussein, and Meschede}}]{schmidt'96}
\bibinfo{author}{\bibfnamefont{O.}~\bibnamefont{Schmidt}},
  \bibinfo{author}{\bibfnamefont{R.}~\bibnamefont{Wynands}},
  \bibinfo{author}{\bibfnamefont{Z.}~\bibnamefont{Hussein}}, \bibnamefont{and}
  \bibinfo{author}{\bibfnamefont{D.}~\bibnamefont{Meschede}},
  \bibinfo{journal}{Phys.\ Rev.\ A} \textbf{\bibinfo{volume}{53}},
  \bibinfo{pages}{R27 } (\bibinfo{year}{1996}).

\bibitem[{\citenamefont{Renzoni and Arimondo}(2000)}]{renzoni'00}
\bibinfo{author}{\bibfnamefont{F.}~\bibnamefont{Renzoni}} \bibnamefont{and}
  \bibinfo{author}{\bibfnamefont{E.}~\bibnamefont{Arimondo}},
  \bibinfo{journal}{Opt.\ Commun.} \textbf{\bibinfo{volume}{178}},
  \bibinfo{pages}{345 } (\bibinfo{year}{2000}).

\bibitem[{\citenamefont{Arimondo}(1996{\natexlab{b}})}]{arimondo'96pra}
\bibinfo{author}{\bibfnamefont{E.}~\bibnamefont{Arimondo}},
  \bibinfo{journal}{Phys.\ Rev.\ A} \textbf{\bibinfo{volume}{54}},
  \bibinfo{pages}{2216 } (\bibinfo{year}{1996}{\natexlab{b}}).

\bibitem[{\citenamefont{Graf et~al.}(1995)\citenamefont{Graf, Arimondo, Fry,
  Nikonov, Padmabandu, Scully, and Zhu}}]{graf'95}
\bibinfo{author}{\bibfnamefont{M.}~\bibnamefont{Graf}},
  \bibinfo{author}{\bibfnamefont{E.}~\bibnamefont{Arimondo}},
  \bibinfo{author}{\bibfnamefont{E.~S.} \bibnamefont{Fry}},
  \bibinfo{author}{\bibfnamefont{D.~E.} \bibnamefont{Nikonov}},
  \bibinfo{author}{\bibfnamefont{G.~G.} \bibnamefont{Padmabandu}},
  \bibinfo{author}{\bibfnamefont{M.~O.} \bibnamefont{Scully}},
  \bibnamefont{and} \bibinfo{author}{\bibfnamefont{S.~Y.} \bibnamefont{Zhu}},
  \bibinfo{journal}{Phys.\ Rev.\ A} \textbf{\bibinfo{volume}{51}},
  \bibinfo{pages}{4030} (\bibinfo{year}{1995}).

\bibitem[{\citenamefont{Bernheim}(1965)}]{bernheim_book}
\bibinfo{author}{\bibfnamefont{R.}~\bibnamefont{Bernheim}},
  \emph{\bibinfo{title}{Optical Pumping}} (\bibinfo{publisher}{W.\ A.\
  Benjamin, Inc., New York}, \bibinfo{year}{1965}).

\bibitem[{\citenamefont{Vanier et~al.}(1998)\citenamefont{Vanier, Godone, and
  Levi}}]{vanier98}
\bibinfo{author}{\bibfnamefont{J.}~\bibnamefont{Vanier}},
  \bibinfo{author}{\bibfnamefont{A.}~\bibnamefont{Godone}}, \bibnamefont{and}
  \bibinfo{author}{\bibfnamefont{F.}~\bibnamefont{Levi}},
  \bibinfo{journal}{Phys. Rev. A} \textbf{\bibinfo{volume}{58}},
  \bibinfo{pages}{2345} (\bibinfo{year}{1998}).

\bibitem[{\citenamefont{Happer}(1972{\natexlab{a}})}]{happer'72}
\bibinfo{author}{\bibfnamefont{W.}~\bibnamefont{Happer}},
  \bibinfo{journal}{Rev.\ Mod.\ Phys.} \textbf{\bibinfo{volume}{44}},
  \bibinfo{pages}{169} (\bibinfo{year}{1972}{\natexlab{a}}).

\bibitem[{\citenamefont{Brandt et~al.}(1997)\citenamefont{Brandt, Nagel,
  Wynands, and Meschede}}]{brandt'97}
\bibinfo{author}{\bibfnamefont{S.}~\bibnamefont{Brandt}},
  \bibinfo{author}{\bibfnamefont{A.}~\bibnamefont{Nagel}},
  \bibinfo{author}{\bibfnamefont{R.}~\bibnamefont{Wynands}}, \bibnamefont{and}
  \bibinfo{author}{\bibfnamefont{D.}~\bibnamefont{Meschede}},
  \bibinfo{journal}{Phys. Rev. A} \textbf{\bibinfo{volume}{56}},
  \bibinfo{pages}{R1063} (\bibinfo{year}{1997}).

\bibitem[{\citenamefont{Wynands and Nagel}(1998)}]{wynands'98}
\bibinfo{author}{\bibfnamefont{R.}~\bibnamefont{Wynands}} \bibnamefont{and}
  \bibinfo{author}{\bibfnamefont{A.}~\bibnamefont{Nagel}},
  \bibinfo{journal}{Appl.\ Phys.\ B} \textbf{\bibinfo{volume}{68}},
  \bibinfo{pages}{1 } (\bibinfo{year}{1998}).

\bibitem[{\citenamefont{Erhard et~al.}(2000)\citenamefont{Erhard, Nu{\ss}mann,
  and Helm}}]{helm'00}
\bibinfo{author}{\bibfnamefont{M.}~\bibnamefont{Erhard}},
  \bibinfo{author}{\bibfnamefont{S.}~\bibnamefont{Nu{\ss}mann}},
  \bibnamefont{and} \bibinfo{author}{\bibfnamefont{H.}~\bibnamefont{Helm}},
  \bibinfo{journal}{Phys. Rev. A} \textbf{\bibinfo{volume}{62}},
  \bibinfo{pages}{061802(R)} (\bibinfo{year}{2000}).

\bibitem[{\citenamefont{Erhard and Helm}(2001)}]{helm'01}
\bibinfo{author}{\bibfnamefont{M.}~\bibnamefont{Erhard}} \bibnamefont{and}
  \bibinfo{author}{\bibfnamefont{H.}~\bibnamefont{Helm}},
  \bibinfo{journal}{Phys. Rev. A} \textbf{\bibinfo{volume}{63}},
  \bibinfo{pages}{043813} (\bibinfo{year}{2001}).

\bibitem[{\citenamefont{Mikhailov
  et~al.}(2004{\natexlab{a}})\citenamefont{Mikhailov, Sautenkov, Rostovtsev,
  and Welch}}]{mikhailov04josab}
\bibinfo{author}{\bibfnamefont{E.~E.} \bibnamefont{Mikhailov}},
  \bibinfo{author}{\bibfnamefont{V.~A.} \bibnamefont{Sautenkov}},
  \bibinfo{author}{\bibfnamefont{Y.~V.} \bibnamefont{Rostovtsev}},
  \bibnamefont{and} \bibinfo{author}{\bibfnamefont{G.~R.} \bibnamefont{Welch}},
  \bibinfo{journal}{JOSA B} \textbf{\bibinfo{volume}{21}}, \bibinfo{pages}{425}
  (\bibinfo{year}{2004}{\natexlab{a}}).

\bibitem[{\citenamefont{Mikhailov
  et~al.}(2004{\natexlab{b}})\citenamefont{Mikhailov, Sautenkov, Novikova, and
  Welch}}]{mikhailov04praprep}
\bibinfo{author}{\bibfnamefont{E.~E.} \bibnamefont{Mikhailov}},
  \bibinfo{author}{\bibfnamefont{V.~A.} \bibnamefont{Sautenkov}},
  \bibinfo{author}{\bibfnamefont{I.}~\bibnamefont{Novikova}}, \bibnamefont{and}
  \bibinfo{author}{\bibfnamefont{G.~R.} \bibnamefont{Welch}},
  \bibinfo{journal}{LANL e-Print archive}
  (\bibinfo{year}{2004}{\natexlab{b}}),
  \urlprefix\url{http://arxiv.org/abs/quant-ph/0402120}.

\bibitem[{\citenamefont{Affolderbach
  et~al.}(2002{\natexlab{a}})\citenamefont{Affolderbach, Knappe, Wynands,
  Taichenachev, and Yudin}}]{affolderbach'02}
\bibinfo{author}{\bibfnamefont{C.}~\bibnamefont{Affolderbach}},
  \bibinfo{author}{\bibfnamefont{S.}~\bibnamefont{Knappe}},
  \bibinfo{author}{\bibfnamefont{R.}~\bibnamefont{Wynands}},
  \bibinfo{author}{\bibfnamefont{A.~V.} \bibnamefont{Taichenachev}},
  \bibnamefont{and} \bibinfo{author}{\bibfnamefont{V.~I.} \bibnamefont{Yudin}},
  \bibinfo{journal}{Phys.\ Rev.\ A} \textbf{\bibinfo{volume}{65}},
  \bibinfo{pages}{043810 } (\bibinfo{year}{2002}{\natexlab{a}}).

\bibitem[{\citenamefont{Akulshin et~al.}(1998)\citenamefont{Akulshin, Barreiro,
  and Lezama}}]{akulshin'98}
\bibinfo{author}{\bibfnamefont{A.~M.} \bibnamefont{Akulshin}},
  \bibinfo{author}{\bibfnamefont{S.}~\bibnamefont{Barreiro}}, \bibnamefont{and}
  \bibinfo{author}{\bibfnamefont{A.}~\bibnamefont{Lezama}},
  \bibinfo{journal}{Phys.\ Rev.\ A} \textbf{\bibinfo{volume}{57}},
  \bibinfo{pages}{2996 } (\bibinfo{year}{1998}).

\bibitem[{\citenamefont{Lezama et~al.}(1999)\citenamefont{Lezama, Barreiro, and
  Akulshin}}]{lazema'99}
\bibinfo{author}{\bibfnamefont{A.}~\bibnamefont{Lezama}},
  \bibinfo{author}{\bibfnamefont{S.}~\bibnamefont{Barreiro}}, \bibnamefont{and}
  \bibinfo{author}{\bibfnamefont{A.~M.} \bibnamefont{Akulshin}},
  \bibinfo{journal}{Phys.\ Rev.\ A} \textbf{\bibinfo{volume}{59}},
  \bibinfo{pages}{4732 } (\bibinfo{year}{1999}).

\bibitem[{\citenamefont{Lipsich et~al.}(2000)\citenamefont{Lipsich, Barreiro,
  Akulshin, and Lezama}}]{lipsich'00}
\bibinfo{author}{\bibfnamefont{A.}~\bibnamefont{Lipsich}},
  \bibinfo{author}{\bibfnamefont{S.}~\bibnamefont{Barreiro}},
  \bibinfo{author}{\bibfnamefont{A.~M.} \bibnamefont{Akulshin}},
  \bibnamefont{and} \bibinfo{author}{\bibfnamefont{A.}~\bibnamefont{Lezama}},
  \bibinfo{journal}{Phys.\ Rev.\ A} \textbf{\bibinfo{volume}{61}},
  \bibinfo{pages}{053803 } (\bibinfo{year}{2000}).

\bibitem[{\citenamefont{Taichenachev et~al.}(1999)\citenamefont{Taichenachev,
  Tumaikin, and Yudin}}]{taichenachev'00jetp}
\bibinfo{author}{\bibfnamefont{A.~V.} \bibnamefont{Taichenachev}},
  \bibinfo{author}{\bibfnamefont{A.~M.} \bibnamefont{Tumaikin}},
  \bibnamefont{and} \bibinfo{author}{\bibfnamefont{V.~I.} \bibnamefont{Yudin}},
  \bibinfo{journal}{JETP Lett.} \textbf{\bibinfo{volume}{69}},
  \bibinfo{pages}{819 } (\bibinfo{year}{1999}).

\bibitem[{\citenamefont{Taichenachev et~al.}(2000)\citenamefont{Taichenachev,
  Tumaikin, and Yudin}}]{taichenachev'00pra}
\bibinfo{author}{\bibfnamefont{A.~V.} \bibnamefont{Taichenachev}},
  \bibinfo{author}{\bibfnamefont{A.~M.} \bibnamefont{Tumaikin}},
  \bibnamefont{and} \bibinfo{author}{\bibfnamefont{V.~I.} \bibnamefont{Yudin}},
  \bibinfo{journal}{Phys.\ Rev.\ A} \textbf{\bibinfo{volume}{61}},
  \bibinfo{pages}{011802 } (\bibinfo{year}{2000}).

\bibitem[{\citenamefont{Kwon et~al.}(2001)\citenamefont{Kwon, Kim, Moon, Park,
  and Kim}}]{kwon'01}
\bibinfo{author}{\bibfnamefont{M.}~\bibnamefont{Kwon}},
  \bibinfo{author}{\bibfnamefont{K.}~\bibnamefont{Kim}},
  \bibinfo{author}{\bibfnamefont{H.~S.} \bibnamefont{Moon}},
  \bibinfo{author}{\bibfnamefont{H.~D.} \bibnamefont{Park}}, \bibnamefont{and}
  \bibinfo{author}{\bibfnamefont{J.~B.} \bibnamefont{Kim}},
  \bibinfo{journal}{J.\ Phys.\ B} \textbf{\bibinfo{volume}{34}},
  \bibinfo{pages}{2951 } (\bibinfo{year}{2001}).

\bibitem[{\citenamefont{Ye et~al.}(2002)\citenamefont{Ye, Rostovtsev, Zibrov,
  and Golubev}}]{ye'02}
\bibinfo{author}{\bibfnamefont{C.~Y.} \bibnamefont{Ye}},
  \bibinfo{author}{\bibfnamefont{Y.~V.} \bibnamefont{Rostovtsev}},
  \bibinfo{author}{\bibfnamefont{A.~S.} \bibnamefont{Zibrov}},
  \bibnamefont{and} \bibinfo{author}{\bibfnamefont{Y.~M.}
  \bibnamefont{Golubev}}, \bibinfo{journal}{Opt.\ Comm.}
  \textbf{\bibinfo{volume}{207}}, \bibinfo{pages}{227 } (\bibinfo{year}{2002}).

\bibitem[{\citenamefont{Dancheva et~al.}(2000)\citenamefont{Dancheva, Alzetta,
  Cartaleva, Taslakov, and Andreeva}}]{dancheva'00}
\bibinfo{author}{\bibfnamefont{Y.}~\bibnamefont{Dancheva}},
  \bibinfo{author}{\bibfnamefont{G.}~\bibnamefont{Alzetta}},
  \bibinfo{author}{\bibfnamefont{S.}~\bibnamefont{Cartaleva}},
  \bibinfo{author}{\bibfnamefont{M.}~\bibnamefont{Taslakov}}, \bibnamefont{and}
  \bibinfo{author}{\bibfnamefont{C.}~\bibnamefont{Andreeva}},
  \bibinfo{journal}{Opt.\ Comm.} \textbf{\bibinfo{volume}{178}},
  \bibinfo{pages}{103 } (\bibinfo{year}{2000}).

\bibitem[{\citenamefont{Alzetta et~al.}(2001)\citenamefont{Alzetta, Cartaleva,
  Dancheva, Andreeva, Gozzini, Botti, and Rossi}}]{alzetta'01}
\bibinfo{author}{\bibfnamefont{G.}~\bibnamefont{Alzetta}},
  \bibinfo{author}{\bibfnamefont{S.}~\bibnamefont{Cartaleva}},
  \bibinfo{author}{\bibfnamefont{Y.}~\bibnamefont{Dancheva}},
  \bibinfo{author}{\bibfnamefont{C.}~\bibnamefont{Andreeva}},
  \bibinfo{author}{\bibfnamefont{S.}~\bibnamefont{Gozzini}},
  \bibinfo{author}{\bibfnamefont{L.}~\bibnamefont{Botti}}, \bibnamefont{and}
  \bibinfo{author}{\bibfnamefont{A.}~\bibnamefont{Rossi}},
  \bibinfo{journal}{J.\ of Opt.\ B} \textbf{\bibinfo{volume}{3}},
  \bibinfo{pages}{181 } (\bibinfo{year}{2001}).

\bibitem[{\citenamefont{Renzoni
  et~al.}(2001{\natexlab{a}})\citenamefont{Renzoni, Zimmermann, Verkerk, and
  Arimondo}}]{arimondo'01job}
\bibinfo{author}{\bibfnamefont{F.}~\bibnamefont{Renzoni}},
  \bibinfo{author}{\bibfnamefont{C.}~\bibnamefont{Zimmermann}},
  \bibinfo{author}{\bibfnamefont{P.}~\bibnamefont{Verkerk}}, \bibnamefont{and}
  \bibinfo{author}{\bibfnamefont{E.}~\bibnamefont{Arimondo}},
  \bibinfo{journal}{J.\ Opt.\ B} \textbf{\bibinfo{volume}{3}},
  \bibinfo{pages}{S7 } (\bibinfo{year}{2001}{\natexlab{a}}).

\bibitem[{\citenamefont{Renzoni
  et~al.}(2001{\natexlab{b}})\citenamefont{Renzoni, Cartaleva, Alzetta, and
  Arimondo}}]{arimondo'01pra}
\bibinfo{author}{\bibfnamefont{F.}~\bibnamefont{Renzoni}},
  \bibinfo{author}{\bibfnamefont{S.}~\bibnamefont{Cartaleva}},
  \bibinfo{author}{\bibfnamefont{G.}~\bibnamefont{Alzetta}}, \bibnamefont{and}
  \bibinfo{author}{\bibfnamefont{E.}~\bibnamefont{Arimondo}},
  \bibinfo{journal}{Phys.\ Rev.\ A} \textbf{\bibinfo{volume}{63}},
  \bibinfo{pages}{065401} (\bibinfo{year}{2001}{\natexlab{b}}).

\bibitem[{\citenamefont{Andreeva et~al.}(2002)\citenamefont{Andreeva,
  Cartaleva, Dancheva, Biancalana, Burchianti, Marinelli, Mariotti, Moi, and
  Nasyrov}}]{andreeva'02}
\bibinfo{author}{\bibfnamefont{C.}~\bibnamefont{Andreeva}},
  \bibinfo{author}{\bibfnamefont{S.}~\bibnamefont{Cartaleva}},
  \bibinfo{author}{\bibfnamefont{Y.}~\bibnamefont{Dancheva}},
  \bibinfo{author}{\bibfnamefont{V.}~\bibnamefont{Biancalana}},
  \bibinfo{author}{\bibfnamefont{A.}~\bibnamefont{Burchianti}},
  \bibinfo{author}{\bibfnamefont{C.}~\bibnamefont{Marinelli}},
  \bibinfo{author}{\bibfnamefont{E.}~\bibnamefont{Mariotti}},
  \bibinfo{author}{\bibfnamefont{L.}~\bibnamefont{Moi}}, \bibnamefont{and}
  \bibinfo{author}{\bibfnamefont{K.}~\bibnamefont{Nasyrov}},
  \bibinfo{journal}{Phys.\ Rev.\ A} \textbf{\bibinfo{volume}{66}},
  \bibinfo{pages}{012502 } (\bibinfo{year}{2002}).

\bibitem[{\citenamefont{Failache et~al.}(2003)\citenamefont{Failache, Valente,
  Ban, Lorent, and Lezama}}]{Failache2003pra}
\bibinfo{author}{\bibfnamefont{H.}~\bibnamefont{Failache}},
  \bibinfo{author}{\bibfnamefont{P.}~\bibnamefont{Valente}},
  \bibinfo{author}{\bibfnamefont{G.}~\bibnamefont{Ban}},
  \bibinfo{author}{\bibfnamefont{V.}~\bibnamefont{Lorent}}, \bibnamefont{and}
  \bibinfo{author}{\bibfnamefont{A.}~\bibnamefont{Lezama}},
  \bibinfo{journal}{Phys. Rev. A} \textbf{\bibinfo{volume}{67}},
  \bibinfo{pages}{043810} (\bibinfo{year}{2003}).

\bibitem[{\citenamefont{Agarwal}(1991)}]{agarwal'91aamop}
\bibinfo{author}{\bibfnamefont{G.~S.} \bibnamefont{Agarwal}},
  \bibinfo{journal}{Advan Atom Mol Opt Phys} \textbf{\bibinfo{volume}{29}},
  \bibinfo{pages}{113} (\bibinfo{year}{1991}).

\bibitem[{\citenamefont{Kash et~al.}(1999)\citenamefont{Kash, Sautenkov,
  Zibrov, Hollberg, Welch, Lukin, Rostovtsev, Fry, and Scully}}]{kash99}
\bibinfo{author}{\bibfnamefont{M.~M.} \bibnamefont{Kash}},
  \bibinfo{author}{\bibfnamefont{V.~A.} \bibnamefont{Sautenkov}},
  \bibinfo{author}{\bibfnamefont{A.~S.} \bibnamefont{Zibrov}},
  \bibinfo{author}{\bibfnamefont{L.}~\bibnamefont{Hollberg}},
  \bibinfo{author}{\bibfnamefont{G.~R.} \bibnamefont{Welch}},
  \bibinfo{author}{\bibfnamefont{M.~D.} \bibnamefont{Lukin}},
  \bibinfo{author}{\bibfnamefont{Y.}~\bibnamefont{Rostovtsev}},
  \bibinfo{author}{\bibfnamefont{E.~S.} \bibnamefont{Fry}}, \bibnamefont{and}
  \bibinfo{author}{\bibfnamefont{M.~O.} \bibnamefont{Scully}},
  \bibinfo{journal}{Phys. Rev. Lett.} \textbf{\bibinfo{volume}{82}},
  \bibinfo{pages}{5229} (\bibinfo{year}{1999}).

\bibitem[{\citenamefont{Mikhailov et~al.}(2002)\citenamefont{Mikhailov,
  Rostovtsev, and Welch}}]{mikhailov2002}
\bibinfo{author}{\bibfnamefont{E.~E.} \bibnamefont{Mikhailov}},
  \bibinfo{author}{\bibfnamefont{Y.}~\bibnamefont{Rostovtsev}},
  \bibnamefont{and} \bibinfo{author}{\bibfnamefont{G.~R.} \bibnamefont{Welch}},
  \bibinfo{journal}{Journal of Modern Optics} \textbf{\bibinfo{volume}{49}},
  \bibinfo{pages}{2535} (\bibinfo{year}{2002}).

\bibitem[{\citenamefont{Javan et~al.}(2002)\citenamefont{Javan, Kocharovskaya,
  Lee, and Scully}}]{javan'02}
\bibinfo{author}{\bibfnamefont{A.}~\bibnamefont{Javan}},
  \bibinfo{author}{\bibfnamefont{O.}~\bibnamefont{Kocharovskaya}},
  \bibinfo{author}{\bibfnamefont{H.}~\bibnamefont{Lee}}, \bibnamefont{and}
  \bibinfo{author}{\bibfnamefont{M.~O.} \bibnamefont{Scully}},
  \bibinfo{journal}{Phys. Rev. A} \textbf{\bibinfo{volume}{66}},
  \bibinfo{pages}{013805} (\bibinfo{year}{2002}).

\bibitem[{\citenamefont{Rostovtsev et~al.}(2002)\citenamefont{Rostovtsev,
  Protsenko, Lee, and Javan}}]{rost'02}
\bibinfo{author}{\bibfnamefont{Y.}~\bibnamefont{Rostovtsev}},
  \bibinfo{author}{\bibfnamefont{I.}~\bibnamefont{Protsenko}},
  \bibinfo{author}{\bibfnamefont{H.}~\bibnamefont{Lee}}, \bibnamefont{and}
  \bibinfo{author}{\bibfnamefont{A.}~\bibnamefont{Javan}}, \bibinfo{journal}{J.
  Mod. Opt.} \textbf{\bibinfo{volume}{49}}, \bibinfo{pages}{2501}
  (\bibinfo{year}{2002}).

\bibitem[{\citenamefont{Kuznetsova et~al.}(2002)\citenamefont{Kuznetsova,
  Kocharovskaya, Hemmer, and Scully}}]{kuznetsova'02}
\bibinfo{author}{\bibfnamefont{E.}~\bibnamefont{Kuznetsova}},
  \bibinfo{author}{\bibfnamefont{O.}~\bibnamefont{Kocharovskaya}},
  \bibinfo{author}{\bibfnamefont{P.}~\bibnamefont{Hemmer}}, \bibnamefont{and}
  \bibinfo{author}{\bibfnamefont{M.~O.} \bibnamefont{Scully}},
  \bibinfo{journal}{Phys. Rev. A} \textbf{\bibinfo{volume}{66}},
  \bibinfo{pages}{063802} (\bibinfo{year}{2002}).

\bibitem[{\citenamefont{Lee et~al.}(2003)\citenamefont{Lee, Rostovtsev, Bednar,
  and Javan}}]{lee'03}
\bibinfo{author}{\bibfnamefont{H.}~\bibnamefont{Lee}},
  \bibinfo{author}{\bibfnamefont{Y.}~\bibnamefont{Rostovtsev}},
  \bibinfo{author}{\bibfnamefont{C.~J.} \bibnamefont{Bednar}},
  \bibnamefont{and} \bibinfo{author}{\bibfnamefont{A.}~\bibnamefont{Javan}},
  \bibinfo{journal}{Appl.\ Phys.\ B} \textbf{\bibinfo{volume}{76}},
  \bibinfo{pages}{33} (\bibinfo{year}{2003}).

\bibitem[{\citenamefont{Levi et~al.}(2000)\citenamefont{Levi, Godone, ans
  S.~Micalizio, and Modugno}}]{Levi2000epjd}
\bibinfo{author}{\bibfnamefont{F.}~\bibnamefont{Levi}},
  \bibinfo{author}{\bibfnamefont{A.}~\bibnamefont{Godone}},
  \bibinfo{author}{\bibfnamefont{J.~V.} \bibnamefont{ans S.~Micalizio}},
  \bibnamefont{and} \bibinfo{author}{\bibfnamefont{G.}~\bibnamefont{Modugno}},
  \bibinfo{journal}{Eur. Phys. J. D} \textbf{\bibinfo{volume}{12}},
  \bibinfo{pages}{53} (\bibinfo{year}{2000}).

\bibitem[{\citenamefont{Knappe et~al.}(2003)\citenamefont{Knappe, Stahler,
  Affolderbach, Taichenachev, Yudin, and Wynands}}]{knappe2003}
\bibinfo{author}{\bibfnamefont{S.}~\bibnamefont{Knappe}},
  \bibinfo{author}{\bibfnamefont{M.}~\bibnamefont{Stahler}},
  \bibinfo{author}{\bibfnamefont{C.}~\bibnamefont{Affolderbach}},
  \bibinfo{author}{\bibfnamefont{A.}~\bibnamefont{Taichenachev}},
  \bibinfo{author}{\bibfnamefont{V.}~\bibnamefont{Yudin}}, \bibnamefont{and}
  \bibinfo{author}{\bibfnamefont{R.}~\bibnamefont{Wynands}},
  \bibinfo{journal}{Appl.\ Phys.\ B} \textbf{\bibinfo{volume}{76}},
  \bibinfo{pages}{57} (\bibinfo{year}{2003}).

\bibitem[{\citenamefont{Taichenachev et~al.}(2003)\citenamefont{Taichenachev,
  Yudin, Wynands, St{\"a}hler, Kitching, and Hollberg}}]{taichen2003}
\bibinfo{author}{\bibfnamefont{A.~V.} \bibnamefont{Taichenachev}},
  \bibinfo{author}{\bibfnamefont{V.~I.} \bibnamefont{Yudin}},
  \bibinfo{author}{\bibfnamefont{R.}~\bibnamefont{Wynands}},
  \bibinfo{author}{\bibfnamefont{M.}~\bibnamefont{St{\"a}hler}},
  \bibinfo{author}{\bibfnamefont{J.}~\bibnamefont{Kitching}}, \bibnamefont{and}
  \bibinfo{author}{\bibfnamefont{L.}~\bibnamefont{Hollberg}},
  \bibinfo{journal}{Phys. Rev. A} \textbf{\bibinfo{volume}{67}},
  \bibinfo{pages}{033810} (\bibinfo{year}{2003}).

\bibitem[{\citenamefont{Li and Xiao}(1995)}]{xiao95}
\bibinfo{author}{\bibfnamefont{Y.}~\bibnamefont{Li}} \bibnamefont{and}
  \bibinfo{author}{\bibfnamefont{M.}~\bibnamefont{Xiao}},
  \bibinfo{journal}{Phys. Rev. A} \textbf{\bibinfo{volume}{51}},
  \bibinfo{pages}{4959} (\bibinfo{year}{1995}).

\bibitem[{\citenamefont{Lounis and Cohen-Tannoudji}(1992)}]{lounis92}
\bibinfo{author}{\bibfnamefont{B.}~\bibnamefont{Lounis}} \bibnamefont{and}
  \bibinfo{author}{\bibfnamefont{C.}~\bibnamefont{Cohen-Tannoudji}},
  \bibinfo{journal}{J. Phys. II} \textbf{\bibinfo{volume}{2}},
  \bibinfo{pages}{579} (\bibinfo{year}{1992}).

\bibitem[{\citenamefont{Ottinger et~al.}(1975)\citenamefont{Ottinger, Scheps,
  York, and Gallagher}}]{ottinger'75}
\bibinfo{author}{\bibfnamefont{C.}~\bibnamefont{Ottinger}},
  \bibinfo{author}{\bibfnamefont{R.}~\bibnamefont{Scheps}},
  \bibinfo{author}{\bibfnamefont{G.~W.} \bibnamefont{York}}, \bibnamefont{and}
  \bibinfo{author}{\bibfnamefont{A.}~\bibnamefont{Gallagher}},
  \bibinfo{journal}{Phys.\ Rev.\ A} \textbf{\bibinfo{volume}{11}},
  \bibinfo{pages}{1815 } (\bibinfo{year}{1975}).

\bibitem[{\citenamefont{Vanier and Audoin}(1989)}]{vanier_book}
\bibinfo{author}{\bibfnamefont{J.}~\bibnamefont{Vanier}} \bibnamefont{and}
  \bibinfo{author}{\bibfnamefont{C.}~\bibnamefont{Audoin}},
  \emph{\bibinfo{title}{The Quantum Phisics of Atomic Frequency Standards}},
  vol.~\bibinfo{volume}{1} (\bibinfo{publisher}{Adam Hilger; Philadelphia},
  \bibinfo{year}{1989}).

\bibitem[{\citenamefont{Happer}(1972{\natexlab{b}})}]{happer72}
\bibinfo{author}{\bibfnamefont{W.}~\bibnamefont{Happer}},
  \bibinfo{journal}{Rev. Mod. Phys.} \textbf{\bibinfo{volume}{44}},
  \bibinfo{pages}{169} (\bibinfo{year}{1972}{\natexlab{b}}).

\bibitem[{\citenamefont{Lukin et~al.}(1997{\natexlab{a}})\citenamefont{Lukin,
  Fleischhauer, Zibrov, Robinson, Velichansky, Hollberg, and
  Scully}}]{lukin'97prl}
\bibinfo{author}{\bibfnamefont{M.~D.} \bibnamefont{Lukin}},
  \bibinfo{author}{\bibfnamefont{M.}~\bibnamefont{Fleischhauer}},
  \bibinfo{author}{\bibfnamefont{A.~S.} \bibnamefont{Zibrov}},
  \bibinfo{author}{\bibfnamefont{H.~G.} \bibnamefont{Robinson}},
  \bibinfo{author}{\bibfnamefont{V.~L.} \bibnamefont{Velichansky}},
  \bibinfo{author}{\bibfnamefont{L.}~\bibnamefont{Hollberg}}, \bibnamefont{and}
  \bibinfo{author}{\bibfnamefont{M.~O.} \bibnamefont{Scully}},
  \bibinfo{journal}{Phys. Rev. Lett.} \textbf{\bibinfo{volume}{79}},
  \bibinfo{pages}{2959} (\bibinfo{year}{1997}{\natexlab{a}}).

\bibitem[{\citenamefont{Harris and Hau}(1999)}]{harris99prl}
\bibinfo{author}{\bibfnamefont{S.~E.} \bibnamefont{Harris}} \bibnamefont{and}
  \bibinfo{author}{\bibfnamefont{L.~V.} \bibnamefont{Hau}},
  \bibinfo{journal}{Phys. Rev. Lett.} \textbf{\bibinfo{volume}{82}},
  \bibinfo{pages}{4611} (\bibinfo{year}{1999}).

\bibitem[{\citenamefont{Johnsson and Fleischhauer}(2002)}]{fleischhauer'02}
\bibinfo{author}{\bibfnamefont{M.~T.} \bibnamefont{Johnsson}} \bibnamefont{and}
  \bibinfo{author}{\bibfnamefont{M.}~\bibnamefont{Fleischhauer}},
  \bibinfo{journal}{Phys. Rev. A} \textbf{\bibinfo{volume}{66}},
  \bibinfo{pages}{043808} (\bibinfo{year}{2002}).

\bibitem[{\citenamefont{Affolderbach
  et~al.}(2002{\natexlab{b}})\citenamefont{Affolderbach, Stahler, Knappe, and
  Wynands}}]{wynands'02}
\bibinfo{author}{\bibfnamefont{C.}~\bibnamefont{Affolderbach}},
  \bibinfo{author}{\bibfnamefont{M.}~\bibnamefont{Stahler}},
  \bibinfo{author}{\bibfnamefont{S.}~\bibnamefont{Knappe}}, \bibnamefont{and}
  \bibinfo{author}{\bibfnamefont{R.}~\bibnamefont{Wynands}},
  \bibinfo{journal}{Appl. Phys. B} \textbf{\bibinfo{volume}{75}},
  \bibinfo{pages}{605} (\bibinfo{year}{2002}{\natexlab{b}}).

\bibitem[{\citenamefont{Budker et~al.}(2000)\citenamefont{Budker, Kimball,
  Rochester, Yashchuk, and Zolotorev}}]{budker'00}
\bibinfo{author}{\bibfnamefont{D.}~\bibnamefont{Budker}},
  \bibinfo{author}{\bibfnamefont{D.~F.} \bibnamefont{Kimball}},
  \bibinfo{author}{\bibfnamefont{S.~M.} \bibnamefont{Rochester}},
  \bibinfo{author}{\bibfnamefont{V.~V.} \bibnamefont{Yashchuk}},
  \bibnamefont{and}
  \bibinfo{author}{\bibfnamefont{M.}~\bibnamefont{Zolotorev}},
  \bibinfo{journal}{Phys. Rev. A} \textbf{\bibinfo{volume}{62}},
  \bibinfo{pages}{043403} (\bibinfo{year}{2000}).

\bibitem[{\citenamefont{Kitching et~al.}(2002)\citenamefont{Kitching, Knappe,
  and Hollberg}}]{hollberg'02}
\bibinfo{author}{\bibfnamefont{J.}~\bibnamefont{Kitching}},
  \bibinfo{author}{\bibfnamefont{S.}~\bibnamefont{Knappe}}, \bibnamefont{and}
  \bibinfo{author}{\bibfnamefont{L.}~\bibnamefont{Hollberg}},
  \bibinfo{journal}{Appl. Phys. Lett.} \textbf{\bibinfo{volume}{81}},
  \bibinfo{pages}{553} (\bibinfo{year}{2002}).

\bibitem[{\citenamefont{Merimaa et~al.}(2003)\citenamefont{Merimaa, Lindvall,
  Tittonen, and Ikonen}}]{merimaa'03}
\bibinfo{author}{\bibfnamefont{M.}~\bibnamefont{Merimaa}},
  \bibinfo{author}{\bibfnamefont{T.}~\bibnamefont{Lindvall}},
  \bibinfo{author}{\bibfnamefont{I.}~\bibnamefont{Tittonen}}, \bibnamefont{and}
  \bibinfo{author}{\bibfnamefont{E.}~\bibnamefont{Ikonen}},
  \bibinfo{journal}{J. Opt. Soc. Am. B} \textbf{\bibinfo{volume}{20}},
  \bibinfo{pages}{273} (\bibinfo{year}{2003}).

\bibitem[{\citenamefont{Lukin et~al.}(1997{\natexlab{b}})\citenamefont{Lukin,
  Fleischhauer, Zibrov, Robinson, Velichansky, Hollberg, and
  Scully}}]{lukin97prl}
\bibinfo{author}{\bibfnamefont{M.~D.} \bibnamefont{Lukin}},
  \bibinfo{author}{\bibfnamefont{M.}~\bibnamefont{Fleischhauer}},
  \bibinfo{author}{\bibfnamefont{A.~S.} \bibnamefont{Zibrov}},
  \bibinfo{author}{\bibfnamefont{H.~G.} \bibnamefont{Robinson}},
  \bibinfo{author}{\bibfnamefont{V.~L.} \bibnamefont{Velichansky}},
  \bibinfo{author}{\bibfnamefont{L.}~\bibnamefont{Hollberg}}, \bibnamefont{and}
  \bibinfo{author}{\bibfnamefont{M.~O.} \bibnamefont{Scully}},
  \bibinfo{journal}{Phys. Rev. Lett.} \textbf{\bibinfo{volume}{79}},
  \bibinfo{pages}{2959} (\bibinfo{year}{1997}{\natexlab{b}}).

\bibitem[{\citenamefont{Sautenkov et~al.}(1999)\citenamefont{Sautenkov, Kash,
  Velichansky, and Welch}}]{sautenkov99las}
\bibinfo{author}{\bibfnamefont{V.}~\bibnamefont{Sautenkov}},
  \bibinfo{author}{\bibfnamefont{M.}~\bibnamefont{Kash}},
  \bibinfo{author}{\bibfnamefont{V.}~\bibnamefont{Velichansky}},
  \bibnamefont{and} \bibinfo{author}{\bibfnamefont{G.}~\bibnamefont{Welch}},
  \bibinfo{journal}{Laser Physics} \textbf{\bibinfo{volume}{9}},
  \bibinfo{pages}{889} (\bibinfo{year}{1999}).

\bibitem[{\citenamefont{Budker et~al.}(2002)\citenamefont{Budker, Gawlik,
  Kimball, Rochester, Yashchuk, and Weis}}]{budker2002rmp}
\bibinfo{author}{\bibfnamefont{D.}~\bibnamefont{Budker}},
  \bibinfo{author}{\bibfnamefont{W.}~\bibnamefont{Gawlik}},
  \bibinfo{author}{\bibfnamefont{D.}~\bibnamefont{Kimball}},
  \bibinfo{author}{\bibfnamefont{S.}~\bibnamefont{Rochester}},
  \bibinfo{author}{\bibfnamefont{V.}~\bibnamefont{Yashchuk}}, \bibnamefont{and}
  \bibinfo{author}{\bibfnamefont{A.}~\bibnamefont{Weis}},
  \bibinfo{journal}{Rev.\ Mod.\ Phys.} \textbf{\bibinfo{volume}{74}},
  \bibinfo{pages}{1153} (\bibinfo{year}{2002}).

\bibitem[{\citenamefont{Sautenkov et~al.}(2000)\citenamefont{Sautenkov, Lukin,
  Bednar, Novikova, Mikhailov, Fleischhauer, Velichansky, Welch, and
  Scully}}]{sautenkov'00}
\bibinfo{author}{\bibfnamefont{V.~A.} \bibnamefont{Sautenkov}},
  \bibinfo{author}{\bibfnamefont{M.~D.} \bibnamefont{Lukin}},
  \bibinfo{author}{\bibfnamefont{C.~J.} \bibnamefont{Bednar}},
  \bibinfo{author}{\bibfnamefont{I.}~\bibnamefont{Novikova}},
  \bibinfo{author}{\bibfnamefont{E.}~\bibnamefont{Mikhailov}},
  \bibinfo{author}{\bibfnamefont{M.}~\bibnamefont{Fleischhauer}},
  \bibinfo{author}{\bibfnamefont{V.~L.} \bibnamefont{Velichansky}},
  \bibinfo{author}{\bibfnamefont{G.~R.} \bibnamefont{Welch}}, \bibnamefont{and}
  \bibinfo{author}{\bibfnamefont{M.~O.} \bibnamefont{Scully}},
  \bibinfo{journal}{Phys.\ Rev.\ A} \textbf{\bibinfo{volume}{62}},
  \bibinfo{pages}{023810} (\bibinfo{year}{2000}).

\bibitem[{\citenamefont{Novikova et~al.}(2001)\citenamefont{Novikova, Matsko,
  and Welch}}]{novikova'01ol}
\bibinfo{author}{\bibfnamefont{I.}~\bibnamefont{Novikova}},
  \bibinfo{author}{\bibfnamefont{A.~B.} \bibnamefont{Matsko}},
  \bibnamefont{and} \bibinfo{author}{\bibfnamefont{G.~R.} \bibnamefont{Welch}},
  \bibinfo{journal}{Opt. Lett.} \textbf{\bibinfo{volume}{26}},
  \bibinfo{pages}{1016} (\bibinfo{year}{2001}).

\bibitem[{\citenamefont{Th\'{e}obald et~al.}(1989)\citenamefont{Th\'{e}obald,
  Dimarcq, Giordano, and C\'{e}rez}}]{theobald'89}
\bibinfo{author}{\bibfnamefont{G.}~\bibnamefont{Th\'{e}obald}},
  \bibinfo{author}{\bibfnamefont{N.}~\bibnamefont{Dimarcq}},
  \bibinfo{author}{\bibfnamefont{V.}~\bibnamefont{Giordano}}, \bibnamefont{and}
  \bibinfo{author}{\bibfnamefont{P.}~\bibnamefont{C\'{e}rez}},
  \bibinfo{journal}{Opt.\ Comm.} \textbf{\bibinfo{volume}{71}},
  \bibinfo{pages}{256 } (\bibinfo{year}{1989}).

\bibitem[{\citenamefont{Schuh et~al.}(1993)\citenamefont{Schuh, Kanorsky, Weis,
  and Hansch}}]{schuh'93}
\bibinfo{author}{\bibfnamefont{B.}~\bibnamefont{Schuh}},
  \bibinfo{author}{\bibfnamefont{S.~I.} \bibnamefont{Kanorsky}},
  \bibinfo{author}{\bibfnamefont{A.}~\bibnamefont{Weis}}, \bibnamefont{and}
  \bibinfo{author}{\bibfnamefont{T.~W.} \bibnamefont{Hansch}},
  \bibinfo{journal}{Opt.\ Comm.} \textbf{\bibinfo{volume}{100}},
  \bibinfo{pages}{451 } (\bibinfo{year}{1993}).

\bibitem[{\citenamefont{Drake et~al.}(1988)\citenamefont{Drake, Lange, and
  Mlynek}}]{drake'88}
\bibinfo{author}{\bibfnamefont{K.~H.} \bibnamefont{Drake}},
  \bibinfo{author}{\bibfnamefont{W.}~\bibnamefont{Lange}}, \bibnamefont{and}
  \bibinfo{author}{\bibfnamefont{J.}~\bibnamefont{Mlynek}},
  \bibinfo{journal}{Opt.\ Comm.} \textbf{\bibinfo{volume}{66}},
  \bibinfo{pages}{315 } (\bibinfo{year}{1988}).

\bibitem[{\citenamefont{Barkov et~al.}(1989)\citenamefont{Barkov,
  Melik-Pashaev, and Zolotorev}}]{barkov'89}
\bibinfo{author}{\bibfnamefont{L.~M.} \bibnamefont{Barkov}},
  \bibinfo{author}{\bibfnamefont{D.~A.} \bibnamefont{Melik-Pashaev}},
  \bibnamefont{and} \bibinfo{author}{\bibfnamefont{M.~S.}
  \bibnamefont{Zolotorev}}, \bibinfo{journal}{Opt.\ Comm.}
  \textbf{\bibinfo{volume}{70}}, \bibinfo{pages}{467 } (\bibinfo{year}{1989}).

\bibitem[{\citenamefont{Baird et~al.}(1989)\citenamefont{Baird, Irie, and
  Wolfenden}}]{baird'89}
\bibinfo{author}{\bibfnamefont{P.~E.~G.} \bibnamefont{Baird}},
  \bibinfo{author}{\bibfnamefont{M.}~\bibnamefont{Irie}}, \bibnamefont{and}
  \bibinfo{author}{\bibfnamefont{T.~D.} \bibnamefont{Wolfenden}},
  \bibinfo{journal}{J.\ Phys.\ B} \textbf{\bibinfo{volume}{22}},
  \bibinfo{pages}{1733 } (\bibinfo{year}{1989}).

\bibitem[{\citenamefont{Chen et~al.}(1990)\citenamefont{Chen, Telegdi, and
  Weis}}]{chen'90a}
\bibinfo{author}{\bibfnamefont{X.}~\bibnamefont{Chen}},
  \bibinfo{author}{\bibfnamefont{V.~L.} \bibnamefont{Telegdi}},
  \bibnamefont{and} \bibinfo{author}{\bibfnamefont{A.}~\bibnamefont{Weis}},
  \bibinfo{journal}{Opt.\ Comm.} \textbf{\bibinfo{volume}{78}},
  \bibinfo{pages}{337 } (\bibinfo{year}{1990}).

\bibitem[{\citenamefont{Kanorsky et~al.}(1995)\citenamefont{Kanorsky, Weis, and
  Skalla}}]{kanorsky'95}
\bibinfo{author}{\bibfnamefont{S.~I.} \bibnamefont{Kanorsky}},
  \bibinfo{author}{\bibfnamefont{A.}~\bibnamefont{Weis}}, \bibnamefont{and}
  \bibinfo{author}{\bibfnamefont{J.}~\bibnamefont{Skalla}},
  \bibinfo{journal}{Appl.\ Phys.\ B} \textbf{\bibinfo{volume}{60}},
  \bibinfo{pages}{S165 } (\bibinfo{year}{1995}).

\bibitem[{\citenamefont{Budker et~al.}(1998)\citenamefont{Budker, Yashchuk, and
  Zolotorev}}]{budker'98}
\bibinfo{author}{\bibfnamefont{D.}~\bibnamefont{Budker}},
  \bibinfo{author}{\bibfnamefont{V.}~\bibnamefont{Yashchuk}}, \bibnamefont{and}
  \bibinfo{author}{\bibfnamefont{M.}~\bibnamefont{Zolotorev}},
  \bibinfo{journal}{Phys.\ Rev.\ Lett.} \textbf{\bibinfo{volume}{81}},
  \bibinfo{pages}{5788 } (\bibinfo{year}{1998}).

\bibitem[{\citenamefont{Budker et~al.}(1999)\citenamefont{Budker, Orlando, and
  Yashchuk}}]{budker'99ajp}
\bibinfo{author}{\bibfnamefont{D.}~\bibnamefont{Budker}},
  \bibinfo{author}{\bibfnamefont{D.~J.} \bibnamefont{Orlando}},
  \bibnamefont{and} \bibinfo{author}{\bibfnamefont{V.}~\bibnamefont{Yashchuk}},
  \bibinfo{journal}{Am.\ J.\ Phys.} \textbf{\bibinfo{volume}{67}},
  \bibinfo{pages}{584 } (\bibinfo{year}{1999}).

\bibitem[{\citenamefont{Zibrov et~al.}(2001)\citenamefont{Zibrov, Novikova, and
  Matsko}}]{zibrov'01ol}
\bibinfo{author}{\bibfnamefont{A.~S.} \bibnamefont{Zibrov}},
  \bibinfo{author}{\bibfnamefont{I.}~\bibnamefont{Novikova}}, \bibnamefont{and}
  \bibinfo{author}{\bibfnamefont{A.~B.} \bibnamefont{Matsko}},
  \bibinfo{journal}{Opt.\ Lett.} \textbf{\bibinfo{volume}{26}},
  \bibinfo{pages}{1311 } (\bibinfo{year}{2001}).

\bibitem[{\citenamefont{Novikova et~al.}(2002)\citenamefont{Novikova, Matsko,
  and Welch}}]{novikova'02apl}
\bibinfo{author}{\bibfnamefont{I.}~\bibnamefont{Novikova}},
  \bibinfo{author}{\bibfnamefont{A.~B.} \bibnamefont{Matsko}},
  \bibnamefont{and} \bibinfo{author}{\bibfnamefont{G.~R.} \bibnamefont{Welch}},
  \bibinfo{journal}{Appl.\ Phys.\ Lett.} \textbf{\bibinfo{volume}{81}},
  \bibinfo{pages}{193 } (\bibinfo{year}{2002}).

\bibitem[{\citenamefont{Rapol et~al.}(2003)\citenamefont{Rapol, Wasan, and
  Natarajan}}]{rapol03}
\bibinfo{author}{\bibfnamefont{U.~D.} \bibnamefont{Rapol}},
  \bibinfo{author}{\bibfnamefont{A.}~\bibnamefont{Wasan}}, \bibnamefont{and}
  \bibinfo{author}{\bibfnamefont{V.}~\bibnamefont{Natarajan}},
  \bibinfo{journal}{Phys. Rev. A} \textbf{\bibinfo{volume}{67}},
  \bibinfo{pages}{053802} (\bibinfo{year}{2003}).

\end{thebibliography}
\end{document}